\documentclass[11pt]{article}

\usepackage[T1]{fontenc}
\usepackage[utf8]{inputenc}
\usepackage{tikz}
\usepackage{pgfplots}
\pgfplotsset{compat=1.18}
\usepackage{amsthm}
\newtheorem{example}{Example}
\usepackage{amsmath,amssymb,amsthm}
\usepackage[margin=0.9in]{geometry}
\usepackage[hidelinks]{hyperref}

\newtheorem{definition}{Definition}
\newtheorem{theorem}{Theorem}
\newtheorem{proposition}{Proposition}
\newtheorem{lemma}{Lemma}
\newtheorem{corollary}{Corollary}

\usepackage{algorithm}
\usepackage{algpseudocode}
\usepackage{xfrac}
\usepackage{cleveref}
\usepackage{natbib}

\usepackage[suppress]{color-edits}
\addauthor[Fatih]{fk}{magenta}
\addauthor[Evi]{em}{olive}

\title{Proportional Representation for Almost-Solid Coalitions and the Proportional Veto Core for Rankings}

\newcommand{\disag}{D}
\author{ 
   \textbf{Evi Micha} and \textbf{Fatih Erdem Kizilkaya}  
  \\
Viterbi School of Engineering \\
  University of Southern California\\
}

\date{}

\begin{document}
\maketitle

\begin{abstract}
   
A proportional ranking should give groups of voters representation commensurate with their size at every prefix. Proportionality for solid coalitions (PSC) provides this guarantee to groups whose members rank the same set of candidates above all others. How much representation should a group receive when a few outside candidates disrupt that agreement?

We extend the proportional veto core to rankings through candidatewise and setwise guarantees. We show that setwise veto protection is equivalent to robust Rank-PSC under reversal of both the voters' preferences and the collective ranking. Robust Rank-PSC extends representation guarantees to almost-solid coalitions at every prefix. Each distinct outside candidate that disrupts a coalition's agreement reduces its guaranteed representation by at most one, and this bound is tight. Using reverse-sequential variants of Ordered Phragm\'en and  STV, we construct rankings satisfying robust Rank-PSC$+$ in polynomial time. This stronger axiom preserves the guarantee after deleting any suffix of the collective ranking and restricting voters' preferences to the remaining candidates. Verifying robust Rank-PSC is coNP-complete, whereas verifying robust Rank-PSC$+$ takes polynomial time.

We also introduce Rank-Aligned PJR$+$, which requires the first $r$ candidates of the collective ranking to satisfy PJR$+$ when each voter approves their top $r$ candidates. We give a polynomial-time algorithm that meets this requirement simultaneously for every $r$. For fixed-size committees, the Expanding Approvals Rule guarantees Rank-PJR$+$ but can violate robust PSC regardless of how ties are broken or payments are assigned. We give a modification that satisfies both guarantees for any prescribed committee size.

\end{abstract}

\section{Introduction}

Proportional representation in committee selection has received
considerable attention in the field of computational social choice over the past decade
\citep{FaliszewskiEtAl2017}. The central goal is
to ensure that groups of voters influence the selected committee in
proportion to their size. This idea has led to a substantial body
of axioms and algorithms for selecting representative sets of
candidates, particularly in the approval-based setting~\citep{LacknerSkowron2023}. Much less is understood about proportional representation
when voters express their preferences as rankings and the collective outcome is itself a ranking. A committee specifies which candidates are
selected, but a ranking must also decide their relative positions.
Moreover, its prefixes may be used as committees of different
sizes, so their representation guarantees must coexist within a
single nested sequence.

One approach to proportional representation in rankings builds on the seminal notion of proportionality for solid coalitions (PSC), which protects groups of voters who rank the same set of candidates above all remaining candidates \citep{Dummett1984Voting}. Requiring PSC to hold for every prefix of the output ranking amounts to combining proportionality with committee monotonicity, two properties shown to be compatible by \citet{aziz2025committee}. PSC thus provides a natural positional benchmark for proportional rankings. Its guarantees, however, require perfect agreement on the separation between the supported set and the remaining candidates: even a single voter ranking an outside candidate above a member of the supported set prevents the coalition from qualifying as solid for that set, leaving it without any representation guarantee under PSC. This raises
the question of how much representation should survive when
the coalition's agreement is imperfect.

 Another approach treats prefixes of the input rankings as approval sets
and applies proportionality axioms from the approval-based setting. In particular, \citet{BrillPeters2023} introduce Rank-PJR$+$, which extends PJR$+$, a strengthening of proportional justified representation (PJR)~\citep{SanchezFernandezEtAl2017PJR}, by requiring its representation guarantees at every approval depth. Although Rank-PJR$+$ can always be satisfied for a fixed committee size, requiring it simultaneously at every prefix of an output ranking is too demanding, as shown by \citet{aziz2025committee}. Thus,
proportionality guarantees that are individually attainable
for every committee size may become incompatible when the
committees must form the prefixes of a single ranking.

\subsection{Our Approach}

Our goal is to advance the study of proportional representation
in rankings by developing guarantees that accommodate imperfect
agreement and can be satisfied by a complete ordering.
Our starting point is the \emph{proportional veto core},
introduced by \citet{Moulin1981ProportionalVeto}. This principle
gives coalitions the power to exclude candidates in proportion
to their size: a sufficiently large group that unanimously
prefers enough alternatives to a candidate can prevent that
candidate from being selected.

We extend this protection to rankings, where every candidate
must eventually appear and a coalition's objection therefore
constrains how high a candidate can be placed in the output ranking.
We first introduce \emph{Candidatewise Rank-PVC}, which determines
when a coalition can object to an individual candidate appearing
among the first $r$ positions. We then introduce
\emph{Setwise Rank-PVC}, which considers objections to sets
of candidates and constrains how many members of a jointly
opposed set may appear together among the first $r$ positions.
This distinction is important because a coalition may be able to prevent each candidate from appearing too high in the ranking individually, without being able to prevent several candidates it collectively opposes from occupying the same prefix.

Interestingly, reversing both the input preferences and the output ranking reveals a close connection between proportional veto protection and PSC: Setwise Rank-PVC is equivalent to a new representation guarantee for \emph{almost-solid coalitions}. Unlike classical PSC, which requires perfect agreement on a supported set, our guarantee continues to protect coalitions even when this agreement is imperfect. For example, suppose that all members of a coalition rank every candidate in a set above all remaining candidates, except that one voter ranks a single outside candidate above some members of the set. While classical PSC provides no guarantee for this coalition and set, our notion reduces its guaranteed representation by at most one. More generally, the loss is bounded by the number of distinct outside candidates that disrupt the coalition's agreement.%, regardless of how many voters disagree about them. We call this strengthening \emph{robust PSC} and show that the bound is tight: no uniformly smaller loss can be guaranteed in general. We then extend this notion to complete rankings, obtaining \emph{robust Rank-PSC}, which requires the guarantee at every output prefix, and the stronger \emph{robust Rank-PSC$+$}, which preserves the guarantee even after deleting any suffix of the output ranking.

\subsection{Our Results}

More specifically, our contributions are organized as follows.

In~\Cref{sec:PVC-rankings}, we first introduce \emph{Candidatewise Rank-PVC} and show that the classical SerialVeto with a fixed round-robin voter order and SimultaneousVeto procedures, when run once with candidates placed in reverse elimination order, satisfy this guarantee~\citep{IanovskiKondratev2021PVC,KizilkayaKempe2023GeneralizedVeto}. We then demonstrate that candidatewise protection is insufficient to capture a coalition's collective objection to a set of candidates. This motivates our stronger notion of \emph{Setwise Rank-PVC}, which, as we show, is not necessarily satisfied by either single-run construction.

In~\Cref{sec:PSC-rankings},
we introduce \emph{robust PSC}, a strengthening of classical PSC that guarantees representation even when a coalition's agreement on a supported set is imperfect. Specifically, each distinct outside candidate that disrupts this agreement reduces the coalition's guaranteed representation by at most one. We show that this bound is tight: any uniformly smaller penalty would make the guarantee impossible to satisfy in general. We then extend robust PSC to complete rankings, introducing \emph{robust Rank-PSC}, which requires the guarantee at every output prefix, and the stronger \emph{robust Rank-PSC$+$}, which additionally preserves the guarantee after deleting any suffix of the output ranking. Interestingly, in~\Cref{sec:Complexity}, we show that while verifying robust Rank-PSC is coNP-complete, the stronger robust Rank-PSC$+$ admits polynomial-time verification.

In~\Cref{sec:Equivelance}, we establish a fundamental connection between proportional veto protection and representation for almost-solid coalitions. Specifically, we prove that a ranking satisfies Setwise Rank-PVC if and only if reversing both the input preferences and the output ranking yields a ranking satisfying robust Rank-PSC. This equivalence provides a bridge between the two notions, allowing algorithms designed to satisfy one to be used to construct rankings satisfying the other.

In~\Cref{sec:setwise-existence},
we establish polynomial-time algorithms for constructing rankings satisfying robust Rank-PSC$+$. 
Our approach extends the reverse-sequential framework of \citet{aziz2025committee}, showing that recursively applying any committee rule satisfying a weak form of robust PSC yields a ranking satisfying robust Rank-PSC$+$.
We prove that Ordered Phragm\'en~\citep{Janson2016Phragmen}  satisfies this weak condition, while  STV satisfies the stronger, robust PSC guarantee~\citep{Dummett1984Voting}.
In contrast, we show that Expanding Approvals Rule (EAR) \citep{AzizLee2020EAR,BrillPeters2023} can violate robust PSC regardless of its tie-breaking and payment choices.

In~\Cref{sec:rank-aligned}, we turn our attention to how far proportionality guarantees can be extended when prefixes of voters' rankings are treated as approval sets.
We introduce \emph{Rank-Aligned PJR$+$}, a relaxation of Rank-PJR$+$ \citep{BrillPeters2023} that requires the first $r$ candidates of the output ranking to satisfy PJR$+$ with respect to voters' top-$r$ candidates, rather than requiring the guarantee for every combination of approval depth and committee size.
We show that, unlike Rank-PJR$+$, this requirement can be satisfied simultaneously at every prefix, and provide a polynomial-time construction based on a modification of EAR.
However, as we discuss, there is a fundamental tension between approval-based proportionality and solid-coalition representation in rankings, leaving limited room for meaningfully combining the two approaches.

In~\Cref{sec:robust-PSC-rank-PJR}, 
motivated by this incompatibility, we investigate whether the two approaches can be reconciled when the outcome is a committee of a fixed size rather than a complete ranking.
Interestingly, we show that robust PSC and  Rank-PJR$+$ can always be satisfied simultaneously.
To establish this result, we first identify why existing rules fail to provide both guarantees: EAR can violate robust PSC, whereas STV and Reverse-Sequential Ordered Phragm\'en can violate Rank-PJR$+$.
We then develop a novel modification of EAR that bridges the two approaches by reserving voter budget for unmet robust PSC claims while maintaining the approval-based representation guarantees.
Our construction establishes the existence of committees satisfying both properties and yields a polynomial-time algorithm for every fixed committee size $k$, although its running time depends exponentially on $k$.

% \noindent
% \textit{Verification (\Cref{sec:Complexity}.)}
% Finally, we study whether a proposed outcome can be checked
% efficiently.
% Checking robust PSC for a committee, or robust Rank-PSC
% for a complete ranking, is coNP-complete.
% Despite imposing additional requirements, robust
% Rank-PSC$+$ can be verified in polynomial time.
% We give a local characterization that reduces verification
% to at most $m-1$ maximum-flow computations, where $m$ is
% the number of candidates. If the ranking fails the axiom,
% the algorithm returns a coalition and a prefix witnessing
% the violation.

\subsection{Related Work}
\label{sec:related-work}

\paragraph{Proportional veto protection.}
\citet{Moulin1981ProportionalVeto} introduced the proportional
veto principle and established nonemptiness of the core.
\citet{IanovskiKondratev2021PVC} gave a polynomial-time algorithm
for computing the core and an anonymous procedure for selecting
from it through simultaneous vetoes.
\citet{KondratevIanovski2024Veto} subsequently studied
veto-core-consistent voting rules through properties such as
monotonicity, participation, and independence of unanimous losers. Both SerialVeto and SimultaneousVeto select a
winner from the proportional veto core~\citep{IanovskiKondratev2021PVC,
KizilkayaKempe2023GeneralizedVeto}.
A related line connects veto protection with metric distortion.
\citet{KizilkayaKempe2022PluralityVeto} introduced Plurality Veto,
which achieves optimal deterministic metric distortion~$3$,
while \citet{KizilkayaKempe2023GeneralizedVeto} developed
generalized veto cores and their connection to matching-based
voting rules.
\citet{KizilkayaKempe2025ApprovalVeto} further studied the tradeoff
between metric distortion and minority protection through
$k$-Approval Veto.
Our candidatewise analysis concerns the positional protection
provided by the complete elimination orders of veto procedures,
including its dependence on the schedule in which voters act.
For approval ballots, \citet{HalpernEtAl2025ApprovalVeto}
introduce flexible-voter representation, extending veto protection
to both single-winner and multiwinner elections. 

\paragraph{Solid coalitions and their generalizations.}
PSC, introduced by \citet{Dummett1984Voting}, is a central
benchmark for proportional representation under ranked
preferences.
\citet{AzizLee2020EAR} generalize PSC to weak preferences and
introduce EAR, combining proportional representation with
improved candidate-monotonicity properties.
\citet{AzizLee2021OrdinalPB} extend the study of solid-coalition
representation to participatory budgeting with ordinal
preferences, introducing inclusion and comparative PSC.
\citet{AzizLee2022Characterization} characterize the committees
satisfying PSC as the possible outcomes of Minimal Demand rules.
Our robust formulation quantifies imperfect solidity by counting
the distinct outside candidates whose removal would make a
coalition solid for a specified set. Each such candidate reduces
the coalition's guaranteed representation by at most one.

\paragraph{Approval-based proportionality and ranked preferences.}
The approval-based committee literature provides another
foundation for proportionality axioms.
\citet{AzizEtAl2017JR} introduce justified representation and
extended justified representation, and
\citet{SanchezFernandezEtAl2017PJR} introduce proportional
justified representation; see \citet{LacknerSkowron2023}
for an overview.
\citet{BrillPeters2023} develop robust and efficiently verifiable
strengthenings, including PJR$+$, and formulate Rank-PJR$+$
by applying the approval-based requirement at every rank
threshold. They show that EAR satisfies Rank-PJR$+$.
Our Rank-Aligned PJR$+$ restricts attention to the approval
depth matching the output prefix size.
Its incompatibility with PSC for rankings shows that even
this relaxation can conflict with solid-coalition representation.
For a fixed committee size, our reserve-based construction
instead combines robust PSC with Rank-PJR$+$.

\paragraph{Proportional rankings.}
\citet{SkowronEtAl2017Rankings} study proportional rankings
when voters report approval sets, evaluating representation
across prefixes of the output ranking.
\citet{IsraelBrill2021Dynamic} extend this setting to account
for sequential selections that need not follow the current
ranking.
For ranked inputs, \citet{aziz2025committee} establish the
compatibility of PSC with committee monotonicity.
They introduce the Solid Coalition Refinement rule and prove
that reverse-sequential application of a weak-PSC rule yields
rankings satisfying PSC.
Their examples include Reverse-Sequential Ordered Phragm\'en,
building on its weak-PSC guarantee \citep{Janson2016Phragmen}.
Our lifting theorem uses the same construction and extends its
analysis to robust representation and restrictions to output
prefixes.

A different approach measures representation through agreement
on pairwise comparisons.
\citet{LedererPetersWas2024SquaredKemeny} study the Squared
Kemeny rule, which minimizes the weighted sum of squared swap
distances to the input rankings, and establish proportionality
guarantees in terms of their weights.
\citet{Lederer2025RankAggregation} develops this approach further
by constructing rules with proportional agreement guarantees
for individual input rankings and groups of rankings.
Our guarantees concern candidate positions and representation
within prefixes of the output ranking.

\section{Preliminaries}

Let \(N\) be a set of \(n\) voters and \(C\) a set of \(m\)
candidates. Each voter \(i\in N\) reports a strict preference
order \(\succ_i\) over \(C\), where \(a\succ_i b\) indicates
that voter \(i\) prefers candidate \(a\) to candidate \(b\).
We denote the resulting preference profile by
\(P=(\succ_i)_{i\in N}\).
For a candidate \(c\in C\), let \(\operatorname{rank}_\tau(c)\)
denote its position  in ranking $\tau$. 
Our main  goal is to aggregate the voters' rankings into a single
social ranking
$\sigma=(c_1,\ldots,c_m)$,
where \(c_r\) is the candidate placed at position \(r\).
For every \(r\in[m]\), let
\(
W_r(\sigma)=\{c_1,\ldots,c_r\}
\)
denote the set of the first \(r\) candidates in the social
ranking \(\sigma\), and let \(W_0(\sigma)=\emptyset\).
In particular, \(c_r\) is the unique candidate added when
moving from \(W_{r-1}(\sigma)\) to \(W_r(\sigma)\).

Although our ultimate goal is to construct a social ranking,
we will also study committee selection. A committee of size
\(k\) is a subset \(W\subseteq C\) with \(|W|=k\).
In particular, each prefix \(W_r(\sigma)\) of a social
ranking can be viewed as a committee of size \(r\).
For every voter $i\in N$ and rank $r\in[m]$,
let $A_i^r$ denote the set of voter $i$'s
top-$r$ candidates according to $\succ_i$.

We consider several established procedures for aggregating
ranked preferences, which we briefly recall for completeness.

\paragraph{Expanding Approvals Rule (EAR).}
Given a target committee size \(k\), EAR~\citep{AzizLee2020EAR} constructs a
committee by gradually expanding voters' approval sets.
Initially, every voter receives a budget \(b_i=k/n\),
and every candidate has a price of \(1\). The procedure
considers approval depths \(r=1,\ldots,m\) in increasing
order. At depth \(r\), voter \(i\) approves the candidates
in \(A_i^r\). Whenever an unelected candidate \(c\) has
total remaining budget of at least \(1\) among its
approvers, it is elected. The approvers contribute exactly
one unit in total, without any voter spending more than
their remaining budget. Their budgets are then updated
accordingly. The procedure continues selecting affordable
candidates at the current depth before moving to the next
one. It terminates when \(k\) candidates have been elected.
Different choices of affordable candidates and payments
may lead to different outcomes. See
\cite{AzizLee2020EAR,BrillPeters2023}.

\paragraph{Ordered Phragm\'en.}
Ordered Phragm\'en constructs a social ranking through a
continuous consumption process. Initially, all candidates
are available. Each voter continuously consumes their
most-preferred available candidate at unit speed, and
each candidate has unit capacity. Once the total
consumption of a candidate reaches one, that candidate is
placed next in the social ranking and is removed from the
set of available candidates. Voters then continue consuming
their most-preferred remaining candidates. The process
continues until all candidates have been ranked. If multiple
candidates become fully consumed simultaneously, ties are
broken arbitrarily. This is also known as Phragm\'en's
ordered method; see \cite{Janson2016Phragmen}.

\paragraph{SerialVeto and SimultaneousVeto.}
We recall two procedures for selecting candidates from the
proportional veto core: SerialVeto and SimultaneousVeto
\citep{IanovskiKondratev2021PVC,KizilkayaKempe2023GeneralizedVeto}.

\emph{SerialVeto.}
Initially, every candidate has a capacity of \(n\).
The voters perform veto operations according to a
prespecified order in which each voter appears exactly
\(m\) times. At each operation, the corresponding voter
decreases by one the capacity of their least-preferred
remaining candidate. Whenever a candidate's capacity
reaches zero, that candidate is eliminated, and subsequent
veto operations are directed only at the remaining
candidates. The procedure continues until all candidates
have been eliminated.

\emph{SimultaneousVeto.}
Initially, every candidate has a capacity of \(n\).
All voters simultaneously veto their least-preferred
remaining candidate at unit speed. Whenever a candidate's
capacity reaches zero, it is eliminated, and the voters
who were vetoing it switch to their least-preferred
remaining candidate. The procedure continues until
all candidates have been eliminated, with ties between
simultaneously eliminated candidates broken arbitrarily. Notice that SimultaneousVeto is equivalent to Ordered
Phragm\'en applied to the reversed preference profile, as noted by~\citep{Peters2023MetricDistortion}.

\section{Generalizing the Proportional Veto Core to Rankings}\label{sec:PVC-rankings}

Our first step is to generalize the Proportional
Veto Core (PVC) from single-winner selection to complete
rankings. We begin by recalling its classical definition.

For a candidate \(a\in C\) and a nonempty coalition
\(T\subseteq N\), define the \emph{joint lower contour}
of \(a\) with respect to \(T\) as
\[
N_a(T)
=
\bigcup_{i\in T}
\left(\{a\}\cup\{c\in C:a\succ_i c\}\right).
\]
Thus, \(N_a(T)\) contains \(a\) and every candidate
that at least one voter in \(T\) ranks below \(a\).
Equivalently, every voter in \(T\) ranks every candidate
outside \(N_a(T)\) strictly above \(a\). Hence, the
smaller \(N_a(T)\) is, the stronger the coalition's
common objection to \(a\).

\begin{definition}[Proportional Veto Core (PVC)]
A candidate \(a\in C\) belongs to the \emph{proportional
veto core} if, for every nonempty coalition \(T\subseteq N\),
\[
\frac{|N_a(T)|}{m}\geq\frac{|T|}{n}.
\]
Equivalently, a coalition \(T\) blocks \(a\) if
\[
\frac{|N_a(T)|}{m}<\frac{|T|}{n}.
\]
\end{definition}

To interpret this condition, consider a coalition \(T\)
containing an \(\alpha=|T|/n\) fraction of the voters.
The coalition can block a candidate if its joint lower
contour contains fewer than an \(\alpha\)-fraction of
the candidates. \cite{Moulin1981ProportionalVeto} showed that the
proportional veto core is always nonempty.

We now ask how this single-winner guarantee can be
extended to complete rankings. In particular, a candidate
placed lower in the ranking need not satisfy the same
proportionality requirement as the first-ranked candidate,
since several candidates have already been placed above it.

We first introduce a candidatewise extension of PVC,
which determines when a coalition can object to a
candidate appearing within a given prefix of the ranking.
We then strengthen this requirement by considering sets
of candidates that appear together within a prefix.

\paragraph{Candidatewise Rank-PVC.}

We begin with a rank-dependent relaxation of the classical
PVC. The idea is that the proportionality requirement
should become weaker as a candidate appears further down
the ranking. More specifically, we scale the original
PVC guarantee by
\(
\sfrac{(m-r+1)}{m},
\)
where \(r\) is the candidate's position in the ranking.
This factor has a natural interpretation in terms of
elimination procedures: a candidate placed at position
\(r\) is the \((m-r+1)\)-st candidate to be eliminated
when candidates are removed from worst to best.

\begin{definition}[Candidatewise Rank-PVC]
A ranking \(\sigma\) satisfies \emph{Candidatewise Rank-PVC} if,
for every candidate \(a\in C\) and every nonempty
coalition \(T\subseteq N\),
\[
\frac{|N_a(T)|}{m}
\geq
\frac{|T|}{n}\cdot
\frac{m-\operatorname{rank}_{\sigma}(a)+1}{m}.
\]
Equivalently, a coalition \(T\) blocks candidate \(a\)
at position \(r=\operatorname{rank}_{\sigma}(a)\) if
\[
\frac{|N_a(T)|}{m}
<
\frac{|T|}{n}\cdot\frac{m-r+1}{m}.
\]
\end{definition}

In particular, the first-ranked candidate must satisfy
the classical PVC guarantee, while the requirement
gradually decreases for candidates appearing further
down the ranking. Note that the factor \(\sfrac{(m-r+1)}{m}\) is tight at every position.
Indeed, on a unanimous profile, Candidatewise Rank-PVC forces the
output to coincide with the common ranking, where the inequality
for \(T=N\) holds with equality at every position.

Running SimultaneousVeto once and ranking candidates in reverse
elimination order yields a ranking that satisfies Candidatewise Rank-PVC.
The same construction with SerialVeto satisfies the guarantee under a
fixed round-robin voter order and can fail under other schedules.
See \Cref{app:candidatewise-veto} for the proofs and a counterexample.

\paragraph{Setwise Rank-PVC.}

A coalition may
have a strong objection to a set of candidates even
though each candidate separately satisfies Candidatewise
Rank-PVC. Consider the following example.

\begin{example}\label{ex:candidatewise-vs-setwise}
Consider an election with five voters and four candidates
$C=\{a,b,c,d\}$, with preferences
\[
\begin{array}{c|l}
\text{Voters} & \text{Preference ranking}\\ \hline
1 & a\succ c\succ b\succ d\\
2 & c\succ a\succ d\succ b\\
3,4,5 & b\succ d\succ a\succ c
\end{array}
\]
and consider the output ranking
$\sigma=(b,d,a,c)$. 
This ranking satisfies Candidatewise Rank-PVC. 
Now consider the coalition $T=\{1,2\}$ and the set $X=\{b,d\}$.
Both voters in $T$ rank every candidate outside $X$ above every
candidate in $X$, although they disagree on the relative order of
$b$ and $d$. Their joint lower contours satisfy
\(
N_b(T)=N_d(T)=\{b,d\}.
\)
When $b$ is considered separately, $d$ contributes to
its lower contour; when $d$ is considered separately,
$b$ contributes to its lower contour. Each candidatewise
check therefore finds a lower contour of size two, but
both checks rely on the same pair of candidates.
The collective objection becomes apparent when we
consider $X=\{b,d\}$ as a whole: its joint lower contour
contains no candidate outside $X$. Both candidates are
ranked below $a$ and $c$ by the same coalition comprising
$40\%$ of the voters, yet together they occupy the entire
top-two prefix. %Setwise Rank-PVC accounts for this shared objection, preventing the members of $X$ from providing separate protection to one another.
\end{example}

This motivates a setwise strengthening.
For a nonempty candidate set \(X\subseteq C\) and a
nonempty coalition \(T\subseteq N\), define the
\emph{joint lower contour} of \(X\) with respect to \(T\)
as
\[
N_X(T)=\bigcup_{a\in X}N_a(T).
\]
Thus, \(N_X(T)\) consists of \(X\) together with every candidate
outside \(X\) that at least one voter in \(T\) ranks below some
member of \(X\). % In particular, \(X\subseteq N_X(T)\).

\begin{definition}[Setwise Rank-PVC]
A ranking \(\sigma\) satisfies \emph{Setwise Rank-PVC}
if, for every nonempty coalition \(T\subseteq N\),
every nonempty set \(X\subseteq C\), and every
\(q\in[m]\),
\[
X\subseteq W_q(\sigma)
\quad\Longrightarrow\quad
|N_X(T)|-|X|+1
\geq
\frac{|T|}{n}(m-q+1).
\]
\end{definition}

The term \(|N_X(T)|-|X|\) counts the candidates outside
\(X\) that belong to its joint lower contour. Thus,
the condition requires that, if all candidates in \(X\)
appear among the first \(q\) positions, the coalition's
joint lower contour must contain sufficiently many
candidates outside \(X\), relative to its size and
the position of the prefix.

For singleton sets, Setwise Rank-PVC reduces exactly
to Candidatewise Rank-PVC.  %Indeed, taking \(X=\{a\}\), the
% condition becomes
% \[
% a\in W_q(\sigma)
% \quad\Longrightarrow\quad
% |N_a(T)|
% \geq
% \frac{|T|}{n}(m-q+1).
% \]
Hence, Setwise Rank-PVC strengthens Candidatewise Rank-PVC
by requiring proportionality not only for individual
candidates, but also for sets of candidates appearing
together within a prefix. In \Cref{ex:candidatewise-vs-setwise}, each of \(b\) and \(d\)
counts toward the other's lower contour. Once we count them
together, no outside candidate contributes, and voters \(1\)
and \(2\) can block the pair from occupying the top two positions.

Unlike the Candidatewise Rank-PVC guarantee, Setwise Rank-PVC
is not necessarily satisfied by simply ranking candidates
in the reverse order of their elimination under
SerialVeto or SimultaneousVeto.  Such an example can be found in~\Cref{app:candidatewise-veto}.

In Section~\ref{sec:setwise-existence}, we show that every preference
profile admits a ranking satisfying Setwise Rank-PVC and that such
a ranking can be computed in polynomial time.

\section{Proportionality For (Almost) Solid Coalitions}\label{sec:PSC-rankings}

In the previous section, we introduced Setwise Rank-PVC. Reversing
both the voters' rankings and the output ranking reveals a connection
with proportionality for solid coalitions (PSC). We first define
robust PSC, which extends representation guarantees to coalitions
whose agreement on a common top set is disrupted by outside
candidates. We then show that a ranking satisfies this guarantee
at every prefix if and only if its reverse satisfies Setwise Rank-PVC
for the reversed preference profile.

We begin with proportionality for solid coalitions (PSC), a
classical representation guarantee for committee elections, introduced by~\cite{Dummett1984Voting}.
Throughout, preferences are strict and we use the Droop version
of PSC.\footnote{Strict preferences mean that each voter ranks
the candidates without ties. For a committee of size $k$, the
Droop version uses the coalition-size condition
$|T|>\ell n/(k+1)$, whereas the Hare version uses
$|T|\ge\ell n/k$, where $n$ is the number of voters.}  A nonempty coalition $T\subseteq N$ \emph{solidly supports}
a nonempty candidate set $X\subseteq C$ if every voter in $T$
ranks every member of $X$ above every candidate outside $X$.

\begin{definition}[PSC]
\label{def:committee-droop-psc}
A committee $W\subseteq C$ of size $k$ satisfies \emph{PSC}
if, for every integer $\ell\ge 1$, every nonempty coalition
$T\subseteq N$, and every nonempty set $X\subseteq C$ solidly
supported by $T$,
\[
|T|>\ell\frac{n}{k+1}
\quad\Longrightarrow\quad
|W\cap X|\ge\min\{\ell,|X|\}.
\]
\end{definition}

For rankings, PSC requires that the top-$k$ committee
$W_k(\sigma)$ satisfy PSC for every $k\in\{1,\ldots,m\}$.
At the level of voting rules, this is equivalent to satisfying
PSC together with \emph{committee monotonicity}: increasing
the target committee size adds candidates without removing
previously selected candidates. The existence of rules satisfying
both properties was recently established by
~\cite{aziz2025committee}. 

However, this guarantee requires perfect agreement on the
separation between $X$ and $C\setminus X$. We next allow
this agreement to be disrupted by outside candidates.

\paragraph{Robust PSC for committees.}

Perfect solidity is a demanding requirement. Instead, we seek a quantitative
guarantee for \emph{almost-solid coalitions}, whose members largely
agree on a preferred set but may disagree about some outside
candidates. Our approach makes the cost of this disagreement
explicit: each distinct outside candidate disrupting the agreement
reduces the coalition's representation guarantee by at most one
seat. We also show that this one-seat loss per defect is unavoidable
if the guarantees must hold simultaneously for all coalitions and
candidate sets.

For a nonempty coalition $T\subseteq N$ and a nonempty candidate
set $X\subseteq C$, define the \emph{cohesion-defect set} of $T$ with
respect to $X$ as
\begin{equation}
\disag_T(X)
=
\left\{
c\in C\setminus X:
\exists i\in T,\ \exists x\in X
\text{ such that }c\succ_i x
\right\}.
\label{eq:cohesion-defect}
\end{equation}
Thus, $\disag_T(X)$ contains the outside candidates that at least
one voter in $T$ ranks above some member of $X$. In particular,
$\disag_T(X)=\emptyset$ if and only if $T$ solidly supports $X$.
%\fkreplace{Its cardinality $\delta_T(X)=|\disag_T(X)|$ is the cohesion defect.}{We refer to its cardinality $|\disag_T(X)|$ as the cohesion defect.} \emcomment{In general, I have avoided using  $\delta_T(X)$ and instead I am using  $|\disag_T(X)|$. I think only in the complexity section there is $\delta$ }

\begin{definition}[Robust PSC]
\label{def:robust-committee-psc}
A size-\(k\) committee \(W\subseteq C\) satisfies
\emph{robust PSC} if, for every integer \(\ell\ge 1\),
every nonempty coalition \(T\subseteq N\) with
\(|T|>\ell n/(k+1)\), and every nonempty set \(X\subseteq C\),
\[
|W\cap X|
\ge
\min\bigl\{|X|,\max\{0,\ell-|\disag_T(X)|\}\bigr\}.
\]
\end{definition}

A coalition large enough to claim \(\ell\) seats starts with a
demand for \(\ell\) members of \(X\). Each distinct outside
candidate in \(\disag_T(X)\) reduces this demand by one.
When \(\disag_T(X)=\emptyset\), the coalition is solid for \(X\)
and the guarantee reduces to ordinary  PSC.

The one-seat penalty per outside candidate is tight. For any
\(k\ge 1\), consider \(k+1\) candidates and \(k+1\) voters
whose rankings are the cyclic shifts of a fixed ranking.
The grand coalition \(T=N\) qualifies for \(\ell=k\) seats.
For any size-\(k\) committee \(W\), choose \(a\in W\) and let
\(X=C\setminus\{a\}\). Since some voter ranks \(a\) first,
\(\disag_T(X)=\{a\}\). The committee contains exactly \(k-1\)
of the \(k\) members of \(X\), so any penalty smaller than one
would require more members of \(X\) than \(W\) contains.

\paragraph{Robust PSC for rankings.}
We extend robust PSC to rankings by requiring it to hold for
every prefix of the output ranking.

\begin{definition}[Robust Rank-PSC]
\label{def:robust-droop-psc}
A ranking $\sigma$ satisfies \emph{robust Rank-PSC} if, for every
$k\in\{1,\ldots,m\}$, the committee $W_k(\sigma)$ satisfies robust
PSC. That is, for every integer
$\ell\ge 1$, every nonempty coalition $T\subseteq N$, and every
nonempty set $X\subseteq C$,
\begin{equation}
|T|>\ell\frac{n}{k+1}
\quad\Longrightarrow\quad
X\subseteq W_k(\sigma)
\ \text{or}\
|X\cap W_k(\sigma)|+|\disag_T(X)|\ge\ell.
\label{eq:robust-droop-psc}
\end{equation}
\end{definition}

Robust Rank-PSC counts defects over the full candidate set \(C\).
We strengthen it by requiring the guarantee to hold after any
suffix of the output ranking is deleted and the voters' rankings
are restricted to the remaining candidates. In the profile
restricted to \(W_t(\sigma)\), the defect set of \(T\) with
respect to \(X\subseteq W_t(\sigma)\) is
\(\disag_T(X)\cap W_t(\sigma)\).

\begin{definition}[Robust Rank-PSC\( + \)]
\label{def:hereditary-robust-rank-psc}
A ranking \(\sigma\) satisfies \emph{robust Rank-PSC\( + \)} if,
for every \(1\le k\le t\le m\), every integer \(\ell\ge 1\),
every nonempty coalition \(T\subseteq N\) with
\(|T|>\ell n/(k+1)\), and every nonempty set \(X\subseteq W_t(\sigma)\),
\[
|W_k(\sigma)\cap X|
\ge
\min\bigl\{|X|,\max\{0,\ell-|\disag_T(X)\cap W_t(\sigma)|\}\bigr\}.
\]
\end{definition}

The case \(t=m\) recovers robust Rank-PSC. As we show in~\Cref{sec:Complexity}, checking whether a
given ranking satisfies robust Rank-PSC is coNP-complete, whereas
robust Rank-PSC\(+\) can be verified in polynomial time.

\section{Equivalence of  Robust Rank-PSC and Reverse Setwise Rank-PVC
}\label{sec:Equivelance}

We next establish a connection between Setwise Rank-PVC and
robust Rank-PSC. Setwise Rank-PVC constrains how high a set of
candidates that a coalition collectively opposes can appear
together in a ranking. Robust Rank-PSC guarantees sufficiently
large coalitions representation from sets they nearly solidly
support. We show that these two requirements are equivalent when
both the voters' rankings and the social ranking are reversed.

Let $P^{\mathrm{rev}}$ be the preference profile obtained
by reversing every voter's ranking in $P$, and let
\(\sigma^{\mathrm{rev}}=(c_m,\ldots,c_1)\)
denote the reverse of the  ranking
$\sigma=(c_1,\ldots,c_m)$.
For a nonempty candidate set $X\subseteq C$ and a
nonempty coalition $T\subseteq N$, let
$N_X^{\mathrm{rev}}(T)$ denote the joint lower contour
of $X$ computed in $P^{\mathrm{rev}}$.

% By the definitions of the joint lower contour and
% the cohesion defect,
% \begin{equation}
% N_X^{\mathrm{rev}}(T)
% =
% X\cup\disag_T(X).
% \label{eq:psc-veto-contour-identity}
% \end{equation}
% Moreover, reversing the social ranking exchanges its
% top and bottom positions. In particular,
% \[
% W_{m-k}(\sigma^{\mathrm{rev}})
% =
% C\setminus W_k(\sigma).
% \]
% These observations allow us to relate the two
% proportionality requirements.

\begin{proposition}
\label{prop:robust-rank-psc-reversal}
A ranking $\sigma$ satisfies robust Rank-PSC for a
preference profile $P$ if and only if
$\sigma^{\mathrm{rev}}$ satisfies Setwise Rank-PVC
for $P^{\mathrm{rev}}$.
\end{proposition}

\begin{proof}
We first show that Setwise Rank-PVC implies robust
Rank-PSC under reversal.

Suppose that $\sigma^{\mathrm{rev}}$ satisfies
Setwise Rank-PVC for $P^{\mathrm{rev}}$.
Fix a committee size $k<m$, an integer $\ell\geq1$,
a nonempty coalition $T\subseteq N$, and a nonempty
candidate set $X\subseteq C$ such that
$
|T|>\ell(\sfrac{n}{(k+1)})$.
We show that
\[
X\subseteq W_k(\sigma)
\quad\text{or}\quad
|W_k(\sigma)\cap X|+|\disag_T(X)|\geq\ell.
\]

If $X\subseteq W_k(\sigma)$, the requirement is
immediately satisfied. Otherwise, let
\(
Y=X\setminus W_k(\sigma)
\)
be the set of candidates in $X$ that are not
included in the top-$k$ committee.
Notice that
$Y\subseteq W_{m-k}(\sigma^{\mathrm{rev}})$
and
$|\disag_T(Y)|
\leq
|\disag_T(X)|+|X\cap W_k(\sigma)|$.

Suppose first that $\sigma^{\mathrm{rev}}$ satisfies
Setwise Rank-PVC. Applying its guarantee to $Y$
at prefix $m-k$ gives
\[
\begin{aligned}
\frac{|T|}{n}(k+1)
&\leq |N_Y^{\mathrm{rev}}(T)|-|Y|+1\\
&=|\disag_T(Y)|+1\\
&\leq|\disag_T(X)|+|X\cap W_k(\sigma)|+1.
\end{aligned}
\]
Consequently, for every integer $\ell\geq1$
such that $|T|>\ell n/(k+1)$, we have
\(
\ell<
|\disag_T(X)|+|X\cap W_k(\sigma)|+1.
\)
Therefore,
\(
|X\cap W_k(\sigma)|+|\disag_T(X)|\geq\ell,
\)
establishing robust Rank-PSC.

Conversely, suppose that $\sigma$ satisfies
robust Rank-PSC. Fix a nonempty set
$X\subseteq W_{m-k}(\sigma^{\mathrm{rev}})$.
Then $X\cap W_k(\sigma)=\emptyset$.
If Setwise Rank-PVC were violated, we would have
\[
\frac{|T|}{n}(k+1)>
|N_X^{\mathrm{rev}}(T)|-|X|+1
=
|\disag_T(X)|+1.
\]
Taking $\ell=|\disag_T(X)|+1$ would give
\(
|T|>\ell(\sfrac{n}{(k+1)}),
\)
while
\(
|X\cap W_k(\sigma)|+|\disag_T(X)|
=\ell-1<\ell,
\)
contradicting robust Rank-PSC.
\end{proof}

\section{Efficient Construction of Robust Rank-PSC$+$ Rankings}\label{sec:setwise-existence}
We next show that robust Rank-PSC$+$ is always satisfiable and
identify a family of rules that satisfy it. Starting with a committee
rule satisfying weak robust PSC, we apply it recursively to construct
the nested prefixes of a ranking. This follows the reverse-sequential
approach of~\cite{aziz2025committee} for ordinary PSC. Since robust Rank-PSC$+$ implies robust Rank-PSC, the equivalence under reversal also gives rules satisfying Setwise Rank-PVC.

\begin{definition}[Weak Robust PSC]
    A committee $W\subseteq C$ of size $k$ satisfies \emph{weak
    robust PSC} if, for every coalition $T\subseteq N$ and every
    nonempty set $X\subseteq C$ with 
    \[
        |T|>
        \bigl(|X|+|\disag_T(X)|\bigr)\frac{n}{k+1}, 
    \]
   then $X\subseteq W$. 
\end{definition}
This is precisely the restriction of robust PSC to demands with $\ell=|X|+|\disag_T(X)|$.
When $|\disag_T(X)|=0$, the condition reduces to weak PSC~\citep{aziz2025committee}.

Let $f$ be a committee rule defined for every target size. Its
\emph{reverse sequentialization} $f^{\mathrm{RS}}$ constructs nested sets
\[
    W_1\subset W_2\subset\cdots\subset W_m=C
\]
backward. Starting from $W_m=C$, at step $t$ it applies $f$ with target size
$t-1$ to the profile restricted to $W_t$ and sets the returned committee equal
to $W_{t-1}$. The resulting ranking is the unique ranking whose top-$k$ set is
$W_k$ for every $k$.

\begin{theorem}[Robust Reverse-Sequential Lifting]
\label{thm:robust-reverse-sequential-lifting}
Let $f$ be a committee rule satisfying
weak robust PSC. Then $f^{\mathrm{RS}}$  satisfies robust Rank-PSC$+$.
\end{theorem}

\begin{proof}
Committee monotonicity follows directly from the nested
construction. We first prove robust Rank-PSC.

Fix a strict preference profile $P$, a target size $k$, an
integer $\ell\ge 1$, a nonempty coalition $T$, and a nonempty
set $X\subseteq C$ such that $|T|>\ell n/(k+1)$.
It suffices to show that
\[
|W_k\cap X|
\ge
\min\{|X|,\ell-|\disag_T(X)|\}.
\]
If $|\disag_T(X)|\ge\ell$, the claim is immediate.
Hence assume that $|\disag_T(X)|<\ell$.

We prove the same bound for $|W_t\cap X|$ by backward induction
on $t=m,m-1,\ldots,k$. The claim holds trivially for $t=m$,
since $W_m=C$. Suppose it holds for some $t\ge k+1$.
If
\[
|W_t\cap X|>\min\{|X|,\ell-|\disag_T(X)|\},
\]
then deleting one candidate from $W_t$ still leaves at least
the required number of candidates from $X$. Thus, it remains
to consider the case
\[
|W_t\cap X|=\min\{|X|,\ell-|\disag_T(X)|\}.
\]

Let $Y=W_t\cap X$, which is nonempty. Since $Y=X\cap W_t$,
we have
\[
\disag_T(Y)\cap W_t\subseteq\disag_T(X).
\]
It follows that
\[
|Y|+|\disag_T(Y)\cap W_t|
\le
\min\{|X|,\ell-|\disag_T(X)|\}+|\disag_T(X)|
\le\ell.
\]

At this step, the rule $f$ is applied to the profile restricted
to $W_t$, with target committee size $t-1$. Its Droop
denominator is therefore $t$. Since $t\ge k+1$, our assumption
on the coalition size implies
\[
|T|>\ell\frac{n}{k+1}
\ge\ell\frac{n}{t}
\ge
\bigl(|Y|+|\disag_T(Y)\cap W_t|\bigr)\frac{n}{t}.
\]
Thus, weak robust PSC applied to the restricted profile
requires $Y\subseteq W_{t-1}$. Consequently,
\[
|W_{t-1}\cap X|
\ge
\min\{|X|,\ell-|\disag_T(X)|\},
\]
as desired.

Finally, fix any \(t\in\{1,\ldots,m\}\) and restrict each voter's
ranking to \(W_t\). The sequence \(W_t,W_{t-1},\ldots,W_1\)
is a valid execution of the reverse-sequential construction on
this restricted profile. By the preceding argument, the induced
ranking on \(W_t\) satisfies robust Rank-PSC in that profile.
Since this holds for every \(t\), the output ranking satisfies
robust Rank-PSC\( + \).
\end{proof}

Every committee rule satisfying weak robust PSC can serve as the
base rule for this construction, yielding a ranking rule that
satisfies robust Rank-PSC$+$. By the reversal equivalence, applying
any such ranking rule to the reversed preference profile and
reversing its output yields a ranking satisfying Setwise Rank-PVC.

We next show that Ordered Phragmén satisfies weak robust PSC and
that STV satisfies robust PSC. We also give an example in which
EAR fails robust PSC.

\subsection{Ordered Phragmén Rule}
\begin{theorem}
\label{thm:ordered-phragmen-weak-robust-psc}
Ordered Phragmén satisfies weak robust PSC.
\end{theorem}

\begin{proof}
Suppose, towards a contradiction, that there exist a nonempty coalition
\(T\subseteq N\) and a nonempty candidate set \(X\subseteq C\) such that
\[
|T|>
\bigl(|X|+|\disag_T(X)|\bigr)\frac{n}{k+1},
\]
but \(X\nsubseteq W_k\), where \(W_k\) is the committee returned by
Ordered Phragmén.

Choose
$x^\star\in X\setminus W_k$,
and let \(\tau\) denote the time at which the \(k\)-th candidate
is fully eaten. We first observe that, at every time before \(\tau\), every voter
in \(T\) eats only candidates in \(X\cup\disag_T(X)\). Indeed,
consider any voter \(i\in T\) and any candidate
\(c\notin X\cup\disag_T(X)\). By the definition of
\(\disag_T(X)\), every candidate in \(X\) is ranked above \(c\)
by voter \(i\).
Since \(x^\star\) is not among the first \(k\) candidates to be
fully eaten, it remains available until time \(\tau\). Hence,
voter \(i\) cannot eat \(c\) before time \(\tau\).

Since the voters in \(T\) jointly eat at rate \(|T|\), they consume
a total mass of \(|T|\tau\) by time \(\tau\), all from candidates
in \(X\cup\disag_T(X)\). The total mass of these candidates is $|X|+|\disag_T(X)|$. (The candidate $x^\star$ may finish exactly at $\tau$ and lose the tie-break.) Thus,
\(
|T|\tau\leq |X|+|D_T(X)|.
\)

On the other hand, among the first \(k\) candidates to be fully eaten, at most
$|X|+|D_T(X)|-1$
belong to \(X\cup D_T(X)\), since \(x^\star\) is not among them. Therefore, at least
$k-|X|-|D_T(X)|+1$
of the first \(k\) candidates lie outside \(X\cup D_T(X)\).

As shown above, voters in \(T\) do not eat any candidate outside \(X\cup D_T(X)\) before time \(\tau\). Hence, all these candidates must be completely eaten by voters in \(N\setminus T\). Since these voters jointly eat at rate \(n-|T|\), we obtain
\[
(n-|T|)\tau
\geq k-|X|-|D_T(X)|+1.
\tag{2}
\]

Substituting the upper bound
\(\tau\le (|X|+|D_T(X)|)/|T|\) into~(2) gives
\[
k-|X|-|D_T(X)|+1
\leq
(n-|T|)\frac{|X|+|D_T(X)|}{|T|}.
\]
Multiplying by \(|T|\) and rearranging yields
$
|T|(k+1)
\leq
n\bigl(|X|+|D_T(X)|\bigr)$
which contradicts the assumed lower bound on \(|T|\). 

\end{proof}
We immediately get the following Corollary. 

\begin{corollary}
\label{cor:rs-ordered-phragmen}
 Reverse-Sequential Ordered Phragmén satisfies
robust Rank-PSC+.
\end{corollary}

\subsection{Single Transferable Vote}
\label{subsec:stv-robust-psc}

STV provides another base rule for the reverse-sequential
construction. We show that it satisfies full robust PSC,
and therefore also the weak condition required by our
lifting theorem.

We use fractional-transfer STV with the exact Droop threshold
\(q=n/(k+1)\), electing candidates whose weight exceeds \(q\)
and transferring their surplus above \(q\).

\begin{theorem}
\label{thm:stv-robust-droop-psc}
Under strict preferences, STV satisfies robust PSC.
\end{theorem}

\begin{proof}
Let $W$ be the committee returned by STV.
Fix a nonempty coalition $T$, a nonempty candidate set $X$,
and an integer $\ell\ge1$ such that $|T|>\ell q$.
If $|\disag_T(X)|\ge\ell$, the claim is immediate.
Otherwise, set
\[
z=\min\{|X|,\ell-|\disag_T(X)|\}.
\]
It suffices to show that at least $z$ members of $X$ are elected.

Label the vote weight originating from $T$.
Every voter in $T$ ranks every member of $X$ above every
candidate outside $X\cup\disag_T(X)$.
Thus, while a member of $X$ remains continuing, all transferable
labeled weight stays within $X\cup\disag_T(X)$.
Each elected candidate retains at most $q$ units of labeled weight.

We claim that the number of elected and continuing members
of $X$ never falls below $z$.
Suppose otherwise, and consider the first elimination that
would violate this claim.
Immediately before this elimination, exactly $z$ members
of $X$ are elected or continuing, including at least one
continuing member.
Let $u$ and $v$ be the numbers of elected and continuing
candidates, respectively, in $X\cup\disag_T(X)$. Then
\[
u+v\le z+|\disag_T(X)|\le\ell.
\]
At most $uq$ units of labeled weight have been retained.
Hence, the $v$ continuing candidates in this set collectively
hold more than
\[
(\ell-u)q\ge vq
\]
units of labeled weight.
Some continuing candidate therefore has weight strictly
greater than $q$, and must be elected before any elimination.
This is a contradiction.

If STV ends by electing all remaining candidates, the claim
immediately gives $|W\cap X|\ge z$.
Otherwise, exactly $k$ candidates have been elected at quota,
leaving total transferable weight
\[
n-kq=q.
\]
Suppose that $|W\cap X|<z$.
By the claim, a member of $X$ remains continuing at termination,
and hence throughout the process.
Consequently, labeled weight has been retained only by elected
candidates in $X\cup\disag_T(X)$.
The remaining labeled weight is therefore at least
\[
\begin{aligned}
|T|-q\bigl(|W\cap X|+|\disag_T(X)|\bigr)
&>\ell q-q\bigl(z-1+|\disag_T(X)|\bigr)\\
&\ge q,
\end{aligned}
\]
contradicting the total remaining weight $q$.
Thus, $|W\cap X|\ge z$, as required.
\end{proof}

We get the following Corollary.

\begin{corollary}
\label{cor:reverse-sequential-stv-robust-psc-plus}
 Reverse-Sequential STV satisfies
robust Rank-PSC$+$.
\end{corollary}

\subsection{EAR Does Not Satisfy Robust PSC}
\label{subsec:ear-robust-psc}

We next show that EAR
does not  satisfy robust PSC.\footnote{In fact,
the counterexample establishes the stronger result that
EAR violates robust PSC even under the Hare quota.}
This failure occurs regardless of how EAR selects
affordable candidates or distributes their costs
among supporting voters.

\begin{proposition}
\label{prop:ear-not-robust-psc}
EAR does not satisfy robust PSC independently of
tie-breaking and payment choices.
\end{proposition}

\begin{proof}
Consider an election with eight voters and five
candidates $C=\{a,b,c,d,x\}$, with preferences
\[
\begin{array}{c|l}
\text{Voters} & \text{Preference ranking}\\ \hline
1 & x\succ c\succ a\succ b\succ d\\
2,3 & a\succ b\succ x\succ c\succ d\\
4,5,6 & b\succ a\succ x\succ d\succ c\\
7 & c\succ a\succ b\succ d\succ x\\
8 & d\succ b\succ a\succ c\succ x
\end{array}
\]

Let the target committee size be $k=4$.
Each voter initially has budget $k/n=1/2$,
and each candidate costs one.

At approval depth $1$, only $a$ and $b$ are
affordable. EAR therefore elects both candidates
before advancing to the next depth. Voters $2,3$
pay the full cost of $a$, while voters $4,5,6$
jointly pay one unit for $b$. Consequently,
voters $4,5,6$ retain a total budget of $1/2$,
regardless of how their payments are divided.

At depth $2$, candidate $c$ is the only affordable
unelected candidate. Indeed, voters $1$ and $7$
both approve $c$ and each has remaining budget
$1/2$. Thus, EAR elects $c$, exhausting their
budgets.

At depth~3, voters \(1,\ldots,6\) approve \(x\). Voters
\(1,2,3\) have exhausted their budgets, while voters \(4,5,6\)
have a combined remaining budget of \(1/2\). Only voter~\(8\)
approves \(d\), with a remaining budget of \(1/2\). Thus, neither
candidate is affordable.

At depth~4, voters \(4,5,6,7,8\) approve \(d\). Voter~\(7\) has
exhausted their budget, but voters \(4,5,6\) can jointly contribute
\(1/2\), and voter~\(8\) can contribute the remaining \(1/2\).
The approvers of \(x\) still have only \(1/2\) in total. Hence EAR
elects \(d\) and returns \(W=\{a,b,c,d\}\).

Now consider the coalition
$T=\{1,2,3,4,5,6\}$
and the candidate set $X=\{x\}$.
The only candidates outside $X$ that some voter
in $T$ ranks above $x$ are $a$ and $b$.
Consequently,
$D_T(X)=\{a,b\}$.
Taking $\ell=3$, the coalition satisfies
\(
|T|=6>3(\sfrac{n}{(k+1)})
=\sfrac{24}{5}.
\)
However, $x\notin W$, and hence
$|W\cap X|+|D_T(X)|
=0+2<3=\ell$.
Thus, $W$ violates robust PSC. 
\end{proof}

\section{Rank-Aligned PJR$+$}\label{sec:rank-aligned}

In the previous section, we studied proportionality in rankings through
robust extensions of PSC. 

A different approach applies approval-based representation axioms to
prefixes of the voters' rankings. In the approval-based setting, let $A_i$
be voter $i$'s approval set and let $W$ be a committee of size $k$.
Following \citet{BrillPeters2023}, $W$ satisfies \emph{PJR$+$}
if, for every integer $\ell\geq 1$ and every nonempty coalition
$S\subseteq N$ with $|S|\geq \ell n/k$, the existence of a candidate
$c\in(\bigcap_{i\in S} A_i)\setminus W$ implies
$|W\cap(\bigcup_{i\in S} A_i)|\geq \ell$. Thus, a single unelected
candidate approved by all of $S$ can witness underrepresentation.
For strict rankings, $W$ satisfies \emph{Rank-PJR$+$} if it satisfies
PJR$+$ on $(A_i^r)_{i\in N}$ for every $r\in[m]$, with the committee
size $k$ held fixed. \citet{BrillPeters2023} show that the EAR satisfies Rank-PJR$+$. Although Rank-PJR$+$ can be satisfied for each fixed committee size,
the resulting committees cannot always be chosen to form the prefixes
of a single ranking. Indeed, \citet{aziz2025committee} show that committee monotonicity can conflict
with   Rank-PJR$+$.

In this section, we investigate a relaxation of Rank-PJR$+$ that
aligns the approval threshold with the committee size. More
specifically, rather than requiring every committee to satisfy
PJR$+$ for every rank cutoff, we require that the first \(r\)
candidates of a ranking satisfy PJR$+$ with respect to the approval
profile in which each voter approves their top \(r\) candidates.
Thus, the committee size and the approval depth increase together.

\begin{definition}[Rank-Aligned PJR$+$]
A ranking \(\sigma\) satisfies \emph{Rank-Aligned PJR$+$}
if, for every \(r\in[m]\), the committee \(W_r(\sigma)\)
satisfies PJR$+$ with respect to the approval profile
\((A_i^r)_{i\in N}\).
\end{definition}

We show that, unlike the full Rank-PJR$+$ requirement, this
rank-aligned notion is always achievable: every strict preference
profile admits a ranking satisfying Rank-Aligned PJR$+$, and such
a ranking can be computed in polynomial time. 

Rank-Aligned PJR$+$ can be achieved by a simple modification of
EAR, in which credits are released
as the approval depth increases.
Initially, every voter has zero credit.
At each stage $r=1,\ldots,m$, give every voter an additional $1/n$
credit and let them approve their top-$r$ candidates.
% Repeatedly select any unselected candidate whose approvers jointly
% hold at least one credit, charging those approvers one credit in
% total without making any balance negative.
% Proceed to the next stage when no candidate is affordable,
% carrying over all unspent credit.

\begin{theorem}\label{thm:rank-aligned-pjrp}
Every strict preference profile admits a ranking satisfying
Rank-Aligned PJR$+$. Moreover, such a ranking can be computed
in polynomial time.
\end{theorem}

\begin{proof}
Let $F_r$
be the set selected by the end of stage $r$.

By stage $r$, exactly $r$ credits have been distributed.
Since each selected candidate costs one credit, $|F_r|\le r$.
At stage $m$, everyone approves every candidate, and the
remaining total credit is $m-|F_m|$.
Thus the stage cannot end with an unselected candidate,
so $F_m=C$.
Let $\sigma$ list candidates in their selection order.
Then $F_r\subseteq W_r(\sigma)$ for every $r$.

Suppose that $\sigma$ violates Rank-Aligned PJR$+$.
There are a stage $r$, a nonempty coalition $S$, and a candidate
\[
c\in
\left(\bigcap_{i\in S}A_i^r\right)\setminus W_r(\sigma)
\]
such that, writing $U_r(S)=\bigcup_{i\in S}A_i^r$,
\[
|S|\ge
\bigl(|W_r(\sigma)\cap U_r(S)|+1\bigr)\frac nr.
\]
By the end of stage $r$, the voters in $S$ have received
$r|S|/n$ credits.
Every earlier payment by these voters went to a candidate
in $U_r(S)$, because approval sets only expand.
Each selected candidate received one credit in total,
so the remaining credit of $S$ is at least
\[
\frac{r|S|}{n}-|F_r\cap U_r(S)|
\ge
\frac{r|S|}{n}-|W_r(\sigma)\cap U_r(S)|
\ge 1.
\]
Since $c\notin F_r$ and every voter in $S$ approves $c$ at
stage $r$, this candidate was still affordable when that
stage ended, a contradiction.

Finally, affordability can be checked by summing the balances
of a candidate's approvers, and payments can be collected
greedily. 
Hence the ranking can be computed in polynomial time.
\end{proof}

Finally, we show that Rank-Aligned PJR$+$ can conflict with PSC.
In the following example, Rank-Aligned PJR$+$ forces a candidate
into the top two positions,\footnote{Already the weaker requirement
of justified representation (JR) at approval depth two forces this
candidate into the top two. For a size-\(k\) committee, JR requires
that whenever at least \(n/k\) voters share an approved candidate,
at least one voter in the group approves a committee member.}
while PSC requires the top four positions to be occupied by the
four other candidates. 

\begin{example}\label{ex:psc-rank-aligned}
Let $N=\{1,2,3,4\}$ and $C=\{a,b,c,d,x\}$. Each voter ranks a
different candidate from $\{a,b,c,d\}$ first and ranks $x$ second;
the remaining candidates can be ordered arbitrarily.
PSC for the size-$4$ prefix requires all four first choices
to be included: each voter forms a solid coalition of size
$1=n/4$ for their top candidate. Therefore $x$ does not belong
to the size-$4$ prefix, and hence does not belong to the size-$2$
prefix either. At approval depth $2$, each voter approves exactly their top
candidate and $x$. If $x$ is missing from the size-$2$ prefix,
the two voters whose first choices are also missing form a
coalition of size $2=n/2$. They commonly approve the unelected
candidate $x$, and neither has any approved candidate in that
prefix. This violates PJR$+$ at depth $2$. 
\end{example}

\section{Jointly Satisfying Robust PSC and Rank-PJR$+$} \label{sec:robust-PSC-rank-PJR}

We now turn from complete rankings to committees of a fixed size.
As \Cref{ex:psc-rank-aligned} shows, even ordinary PSC and
Rank-Aligned PJR$+$ can impose conflicting requirements at different
prefixes of the same ranking. For a fixed committee size, however,
we can choose the committee independently of the outcomes at other
sizes. This leaves open whether a single committee can satisfy both
robust PSC and Rank-PJR$+$. STV and Reverse-Sequential Ordered Phragmén satisfy robust PSC,
while EAR satisfies Rank-PJR$+$. We first discuss intuitively why each of these
rules can fail the other guarantee. We then introduce a modification
of EAR that satisfies both.

\subsection{Why Existing Rules Fail to Combine the Guarantees}

Although both robust PSC and Rank-PJR$+$ impose
proportionality requirements on coalitions, they capture
different aspects of the voters' preferences. Rank-PJR$+$
protects coalitions that share a candidate within their
original top-$r$ sets, whereas robust PSC protects
coalitions that approximately solidly support a set of
candidates. The distinction helps explain why rules
satisfying one guarantee may fail the other.

Consider first STV and Reverse-Sequential Ordered
Phragm\'en. Both procedures make decisions based on
the candidates remaining in the election. As candidates
are eliminated, lower-ranked alternatives move upward
in the restricted preference orders. Consequently,
a coalition's agreement on a candidate within its
original top-$r$ positions may no longer be reflected
in the rankings considered by the procedure.

For example, consider four voters with preferences
\[
\begin{array}{c|l}
1 & a\succ b\succ c\succ d\succ e\succ f\\
2 & e\succ b\succ c\succ d\succ f\succ a\\
3 & d\succ c\succ b\succ f\succ a\succ e\\
4 & f\succ c\succ b\succ e\succ a\succ d
\end{array}
\]
and a target committee size of $k=2$.
With suitable tie-breaking, STV returns
$\{d,f\}$, while Reverse-Sequential Ordered
Phragm\'en returns $\{a,e\}$.

The first committee violates Rank-PJR$+$ because
voters $1,2$ share candidate $b$ in their original
top-two positions but receive no representative
from $A_1^2\cup A_2^2$. Similarly, the second
committee violates Rank-PJR$+$ because voters
$3,4$ share candidate $c$ in their original
top-two positions but receive no representative
from $A_3^2\cup A_4^2$.
 The difficulty is that these procedures respond
to the relative preferences among the remaining
candidates, rather than preserving representation
claims defined at the voters' original rank
positions. 

EAR behaves differently. It preserves the voters'
original rank positions by gradually expanding
their approval sets. However, once a candidate
belongs to a voter's current approval set, EAR
may use that voter's budget to elect the candidate
even when a more-preferred candidate remains
available.

This is precisely what happens in the EAR
counterexample established in~\Cref{prop:ear-not-robust-psc}. Voter $1$
ranks $x\succ_1 c$, but at approval depth $2$,
both candidates are approved. EAR elects $c$
using voter $1$'s remaining budget, even though
$x$ remains unelected. As a result, the coalition
supporting $x$ loses resources needed to satisfy
its robust-PSC claim.

Thus, the two approaches face different
limitations. STV and Reverse-Sequential Ordered
Phragm\'en preserve preferences among the
remaining candidates but do not preserve
claims tied to original rank positions.
EAR preserves the original rank cutoffs, but
its purchasing decisions do not account for
preference distinctions within those cutoffs.

\subsection{A Reserve-Protected Expanding Approvals Rule}

We now introduce a modification of EAR that simultaneously
satisfies both robust PSC and Rank-PJR$+$.
As discussed above, EAR preserves the voters' original
rank positions, but its purchasing decisions may exhaust
the resources needed to satisfy a coalition's robust-PSC
claim. Our approach addresses this issue by imposing
additional restrictions on how voters' budgets can be spent.

Our algorithm, presented in \Cref{alg:reserve-ear}, adapts EAR by protecting the budgets of coalitions with outstanding robust PSC claims. As in EAR, each voter starts with a fixed budget, and approval sets expand one rank at a time. At each depth, a candidate can be elected if its current approvers can pay for it while leaving every coalition with enough budget to meet its remaining robust PSC claim. The amount reserved for a coalition depends on how much representation it has already received and which outside candidates disrupt its agreement.

%The proof shows that these reserves protect robust PSC claims while allowing the rule to satisfy Rank-PJR$+$ across all approval depths. Once every voter approves every candidate, there is always a purchase that respects the reserves whenever a seat remains unfilled. The rule therefore fills all $k$ seats.

\begin{algorithm}[t] \label{EAR-with-reservations}
\caption{Reserve-Protected Expanding Approvals Rule}
\label{alg:reserve-ear}
\begin{algorithmic}[1]
\Require Profile $P$, committee size $k$
\Ensure A committee $W$ of size $k$

\State $W\gets\emptyset$
\State $b_i\gets 1$ for every voter $i\in N$
\State $r\gets 1$

\While{$|W|<k$}
    \State Find a candidate $c\in C\setminus W$ and payments
    $(p_i)_{i\in N}$ satisfying:
    \Statex \hspace{\algorithmicindent}
    $0\leq p_i\leq b_i$ for every $i\in N$
    \Statex \hspace{\algorithmicindent}
    $\displaystyle\sum_{i\in N}p_i=q$
    \Statex \hspace{\algorithmicindent}
    $p_i=0$ whenever $c\notin A_i^r$
    \Statex \hspace{\algorithmicindent}
    $\displaystyle
    \sum_{i\in T_j}(b_i-p_i)
    \geq R_j(W\cup\{c\})
    \quad\text{for every claim }j\in\mathcal J$

    \If{such a candidate $c$ and payments exist}
        \State $W\gets W\cup\{c\}$
        \ForAll{$i\in N$}
            \State $b_i\gets b_i-p_i$
        \EndFor
    \ElsIf{$r<m$}
        \State $r\gets r+1$
    \Else
        \State \Return Failure
    \EndIf
\EndWhile

\State \Return $W$
\end{algorithmic}
\end{algorithm}

\subsubsection{Description of the Algorithm.} Initially, each voter has one unit of virtual money, and each candidate has a fixed price
\[
q=\frac{n}{k+1}+\frac{1}{(k+1)^2}.
\]
Notice that
\[
\frac{n}{k+1}<q<\frac{n}{k}.
\]
For every integer $1\le\ell\le k$, integrality gives the quota conversion
\[
|T|>\ell\frac{n}{k+1}\quad\Longleftrightarrow\quad |T|\ge\ell q.
\]
Indeed, a positive difference between $|T|$ and $\ell n/(k+1)$ is at least $1/(k+1)$, whereas $\ell/(k+1)^2<1/(k+1)$. The converse follows from $q>n/(k+1)$.

As in EAR, the algorithm considers approval depths \(r=1,\ldots,m\)
in increasing order. At depth \(r\), each voter approves their top
\(r\) candidates. An unelected candidate can be purchased if its
approvers can pay the price \(q\) from their remaining budgets
while preserving the reserves defined below. After each purchase,
the algorithm deducts the payments and continues at the same depth.
It advances to the next depth only when no further purchase is
possible and stops once \(k\) candidates have been elected.

The reserves protect the representation that robust PSC guarantees
to each coalition. Since every voter starts with one unit of budget,
a nonempty coalition \(T\subseteq N\) can initially afford
\[
L(T)=\left\lfloor\frac{|T|}{q}\right\rfloor
\]
candidates. By the quota conversion above, \(L(T)\) is the largest
number of seats that \(T\) can claim, with \(L(T)=0\) if it cannot
claim a seat. For each nonempty candidate set \(X\subseteq C\),
the coalition's entitlement from \(X\) is therefore
\[
h(T,X)=\min\bigl\{|X|,\max\{0,L(T)-|\disag_T(X)|\}\bigr\}.
\]

We call each pair \((T,X)\) a \emph{claim}: it requires the final
committee to contain at least \(h(T,X)\) members of \(X\).
Let
\[
\mathcal J=\{(T,X):\varnothing\ne T\subseteq N,\ 
                     \varnothing\ne X\subseteq C\}
\]
be the set of all claims. A size-\(k\) committee \(W\) satisfies
robust PSC if and only if \(|W\cap X|\ge h(T,X)\) for every
\((T,X)\in\mathcal J\). For a claim \(j=(T_j,X_j)\), write
\(h_j=h(T_j,X_j)\) for its entitlement and
\(D_j=\disag_{T_j}(X_j)\) for its defect set.

The state of the algorithm consists of the partial committee \(W\),
the current approval depth \(r\), and each voter's remaining
budget \(b_i\). For a coalition \(T\), let
\(b(T)=\sum_{i\in T} b_i\) denote its total remaining budget.
A claim \(j=(T_j,X_j)\) is \emph{active} if
\(|W\cap X_j|<h_j\). While the claim is active, the coalition
must retain enough budget to pay for each representative it is
still owed from \(X_j\), plus one candidate price for each
unelected member of its defect set \(D_j\). Once the claim is
fulfilled, its reserve drops to zero. For each claim \(j\), define its reserve by
\[
R_j(W)=
\begin{cases}
q\bigl(h_j-|W\cap X_j|+|D_j\setminus W|\bigr),
    & \text{if } |W\cap X_j|<h_j,\\
0,  & \text{otherwise}.
\end{cases}
\]
The algorithm maintains \(b(T_j)\ge R_j(W)\) for every claim \(j\).

\paragraph{Safe purchases.}
At rank \(r\), an unelected candidate \(c\) is \emph{safely
purchasable} if voters who rank \(c\) among their top \(r\) choices
can together pay \(q\) while leaving each coalition enough budget
to meet its reserve after \(c\) is elected. Formally, there must
exist payments \((p_i)_{i\in N}\) satisfying
\[
\begin{aligned}
0\le p_i &\le b_i && \text{for every } i\in N,\\
p_i &= 0 && \text{if } c\notin A_i^r,\\
\sum_{i\in N}p_i &= q,\\
b(T_j)-p(T_j) &\ge R_j(W\cup\{c\})
    && \text{for every } j\in\mathcal J,
\end{aligned}
\]
where \(p(T)=\sum_{i\in T}p_i\).

The algorithm starts at rank \(r=1\) with an empty committee and
unit budget for each voter. At the current rank, it elects any
safely purchasable candidate \(c\), adds \(c\) to \(W\), and
deducts a feasible payment \(p_i\) from each voter's budget.
It continues at the same rank until \(k\) candidates have been
elected or no further safe purchase is available. In the latter
case, if seats remain, it increases \(r\) by one and continues.

\begin{theorem}\label{thm:PSC-Rank-PJR}
 For every $k\in\{1,\ldots,m\}$, \Cref{alg:reserve-ear} returns a committee of size $k$ that satisfies both robust PSC and rank-PJR$+$.
\end{theorem}

\subsubsection{Proof of~\Cref{thm:PSC-Rank-PJR}}
The theorem follows from the next two lemmas.

\begin{lemma}\label{lem:alg-robust-psc}
\Cref{alg:reserve-ear} returns a committee that satisfies robust PSC. 
\end{lemma}

\begin{proof}

Initially, every active claim \(j\) satisfies
\[
R_j(\varnothing)=q(h_j+|D_j|)\le qL(T_j)\le |T_j|=b(T_j),
\]
where the first inequality follows from the definition of \(h_j\).
Inactive claims have reserve zero. Thus the reserve invariant holds
initially, and every safe purchase preserves it by definition.

Next, we  show that a safe purchase always exists at rank \(m\), provided that fewer than \(k\) candidates have been elected.

Consider an auxiliary elimination process that does not alter the actual committee or the voters' budgets. Initially, all candidates in \(C\setminus W\) are available. Each voter places their remaining budget on their favorite available candidate. If some candidate receives support of at least \(q\), we stop. Otherwise, we delete an arbitrary available candidate, recompute the voters' favorite available candidates, and repeat.

We claim that the process can never delete the last available candidate from \(X\) for any active claim \((T,X)\in\mathcal J\).

To see this, suppose that we are about to delete the last available candidate from \(X\). Every voter in \(T\) must currently support a candidate in \(X\cup\disag_T(X)\): indeed, every voter in \(T\) ranks the remaining available member of \(X\) above every candidate outside this set. Furthermore, at most
\[
1+|\disag_T(X)\setminus W|
\]
candidates in \(X\cup\disag_T(X)\) are still available. But, by the reserve invariant,
\[
\begin{aligned}
b(T)
&\geq R_{T,X}(W)\\
&=q\bigl(h(T,X)-|W\cap X|
       +|\disag_T(X)\setminus W|\bigr)\\
&\geq q\bigl(1+|\disag_T(X)\setminus W|\bigr).
\end{aligned}
\]
The pigeonhole principle therefore implies that some available candidate already receives support of at least \(q\), contradicting the condition for deletion.

The auxiliary process must eventually stop. In particular, it cannot delete its last available candidate, since if only one candidate remains, it receives the entire remaining budget
\[
b(N)=n-q|W|
\geq n-q(k-1)>q.
\]
Consequently, some candidate \(c\) receives support of at least \(q\). Collect exactly \(q\) from its current supporters without exceeding any voter's remaining budget. We claim that these payments satisfy the reserved claims. Consider any active claim \((T,X)\) such that
\[
c\notin X\cup\disag_T(X).
\]
By the preceding argument, at least one candidate from \(X\) is still available. Every voter in \(T\) ranks this candidate above \(c\). Hence, no voter in \(T\) supports \(c\) in the auxiliary process, and therefore \(p(T)=0\).

For each such claim, both the coalition's budget and its reserve
remain unchanged. For every other active claim, electing \(c\)
either fills one of the seats still owed from \(X\) or removes a
candidate from \(D_T(X)\setminus W\). Its reserve therefore falls
by at least \(q\), while the coalition pays at most \(q\).
Fulfilled claims remain fulfilled, so the purchase preserves every
reserve.  Since every voter approves every candidate at rank \(m\), the purchase is safe at rank \(m\). We have thus shown that, whenever fewer than \(k\) candidates have been elected, a safe purchase exists at full rank. 

Let \(W^*\) be the resulting committee. Since every purchase costs exactly \(q\),
$b(N)=n-kq<q$
at termination. Suppose that some claim \((T,X)\in\mathcal J\) were still active. By  the fact that every active claim has reserve at least \(q\), we would have
$b(T)\geq R_{T,X}(W^*)\geq q$.
This contradicts \(b(T)\leq b(N)<q\). Hence, every claim is fulfilled.

\end{proof}

\begin{lemma}
 \Cref{alg:reserve-ear} returns a committee that satisfies rank-PJR$+$.
\end{lemma}
\begin{proof}

Let $W^*$ be the committee that~\Cref{alg:reserve-ear} returns. 
Suppose, for a contradiction, that \(W^*\) violates rank-PJR$+$. Then there exist a coalition \(S\subseteq N\), a rank \(r\in\{1,\ldots,m\}\), and an integer \(\ell\geq1\) such that
\[
|S|>\ell\frac{n}{k+1},
\]
while
\[
\left|W^*\cap\bigcup_{i\in S}A_i^r\right|\leq\ell-1
\qquad\text{and}\qquad
\bigcap_{i\in S}A_i^r\nsubseteq W^*.
\]
Since \(q<n/k\), it follows that \(|S|\ge \ell q\). Moreover, there exists a candidate
\[
c\in\left(\bigcap_{i\in S}A_i^r\right)\setminus W^*.
\]

We first show that the algorithm cannot have completed the election using only purchases at ranks at most \(r\). Otherwise, every payment by a voter in \(S\) would have gone to
a candidate in \(\bigcup_{i\in S} A_i^r\), since approval sets
only expand with rank. Each elected candidate costs \(q\), so at
termination we would have
\[
\begin{aligned}
b(S)
&\geq |S|-q\left|W^*\cap\bigcup_{i\in S}A_i^r\right|\\
&\geq q\ell-q(\ell-1)\\
&=q,
\end{aligned}
\]
This contradicts $b(S)\le b(N)=n-kq<q$.
Hence, the algorithm must have advanced from rank \(r\) to rank \(r+1\). We  construct a safe purchase at rank \(r\), contradicting the fact that no such purchase exists. 

Consider the state \((W,b)\) immediately before this advance, when no safe purchase was available at rank \(r\). Let
\[
F=(W^*\setminus W)\cap
\bigcup_{i\in S}A_i^r.
\]
Then, we obtain
\[
\begin{aligned}
b(S)
&\geq |S|-q\left|W\cap\bigcup_{i\in S}A_i^r\right|\\
&\geq q\ell-q\left(
\left|W^*\cap\bigcup_{i\in S}A_i^r\right|-|F|
\right)\\
&\geq q(|F|+1).
\end{aligned}
\]

% Now consider the menu \(F\cup\{c\}\). This menu contains exactly \(|F|+1\) candidates, all currently unselected. Each voter in \(S\) assigns their remaining budget to their favorite candidate in this menu. Since
% \[
% c\in\bigcap_{i\in S}A_i^r,
% \]
% every voter in \(S\) ranks \(c\) among their top-\(r\) candidates. Consequently, the favorite candidate of each voter in \(S\) in this menu also belongs to their top-\(r\) set: if a voter chooses \(c\), this is immediate, and otherwise they choose a candidate ranked above \(c\).

Suppose that each voter in \(S\) assigns their entire remaining budget to their favorite candidate in \(F\cup\{c\}\). Since the total budget of \(S\) is at least \(q(|F|+1)\), and the menu contains exactly \(|F|+1\) candidates, the pigeonhole principle implies that some candidate \(d\in F\cup\{c\}\) receives at least \(q\) units of budget. Since every voter in \(S\) ranks \(c\) among their top \(r\)
candidates, every voter who chooses \(d\) from \(F\cup\{c\}\)
also ranks \(d\) among their top \(r\) candidates. We can
therefore collect exactly \(q\) from these voters without
exceeding their remaining budgets.
It remains to show that this purchase preserves every reserve.  Consider an active claim \((T,X)\) such that
\[
d\notin X\cup\disag_T(X).
\]

Suppose, towards a contradiction, that some voter \(i\in S\cap T\) contributes to the purchase of \(d\). By the definition of \(\disag_T(X)\), voter \(i\) ranks every candidate in \(X\) above \(d\). Since \(d\in A_i^r\), this implies
\[
X\subseteq A_i^r
\subseteq\bigcup_{j\in S}A_j^r.
\]

On the other hand, the claim \((T,X)\) is currently active, whereas the final committee \(W^*\) satisfies Robust PSC, by ~\Cref{lem:alg-robust-psc}. Hence,
\[
|W\cap X|<h(T,X)\leq |W^*\cap X|.
\]
Since \(W\subseteq W^*\), there must exist a candidate
\[
x\in(W^*\setminus W)\cap X.
\]
By the preceding inclusion, \(x\in F\). But voter \(i\) ranks \(x\) above \(d\), contradicting the fact that \(i\) assigned their budget to their favorite candidate in the menu \(F\cup\{c\}\).

We conclude that no voter in \(S\cap T\) contributes to the purchase of \(d\). 
By the same argument as in ~\Cref{lem:alg-robust-psc}, these payments preserve every
reserve. Hence \(d\) is safely purchasable at rank \(r\).

This contradicts the fact that the algorithm advanced from rank \(r\) to rank \(r+1\) only because no safe purchase was available at rank \(r\). Therefore, \(W^*\) satisfies rank-PJR$+$ as well.
\end{proof}

\paragraph{Running Time.}
Although reserve-protected EAR maintains constraints for exponentially many voter coalitions and candidate sets, these constraints can be checked without explicitly enumerating them. This yields an algorithm whose running time depends exponentially on the committee size $k$, but is polynomial in the number of voters and candidates for every fixed $k$.

\begin{proposition}
\label{prop:running-time}
\Cref{alg:reserve-ear} computes a committee satisfying robust PSC and  Rank-PJR$^+$ in $m^{O(k)}\operatorname{poly}(n,m)$ time.
\end{proposition}

All the omitted proofs are deferred to~\Cref{app:omitted-proofs}.

\section{Verification Complexity}\label{sec:Complexity}

\subsection{Polynomial-Time Verification of Robust Rank-PSC$+$}
\label{subsec:verification-rank-psc-plus}

We next show that Robust Rank-PSC$+$ can be verified in
polynomial time. The key observation is that its defining
conditions admit a local characterization in terms of the last
candidate in each prefix of the ranking.

For every voter \(i\in N\) and every \(t\in[m]\), define
\[
U_i^t=\{c_t\}\cup\{x\in W_t: x\succ_i c_t\} \quad \text{and} \quad
U_T^t=\bigcup_{i\in T}U_i^t.
\]
Thus, \(U_i^t\) consists of \(c_t\) and all candidates in the
first \(t\) positions that voter \(i\) ranks above \(c_t\).
By the definition of the cohesion defect,
\(U_T^t=\{c_t\}\cup
\bigl(D_T(\{c_t\})\cap W_t\bigr).\) We first show the next lemma. 

\begin{lemma}
\label{thm:robust-rank-psc-plus-local}
A ranking \(\sigma\) satisfies Robust Rank-PSC$+$ if and only if
\(
n|U_T^t|\geq t|T|\)
for every $t\in[m]$
and every nonempty $T\subseteq N$.

\end{lemma}

We now use this characterization to obtain a polynomial-time
verification algorithm.

\begin{theorem}
\label{thm:robust-rank-psc-plus-verification}
Robust Rank-PSC$+$ can be verified in polynomial time. Moreover, if the
ranking violates Robust Rank-PSC$+$, the algorithm returns
a prefix and a coalition witnessing the violation.
\end{theorem}

\begin{proof}
By \Cref{thm:robust-rank-psc-plus-local}, it suffices
to verify that
\(
n|U_T^t|\geq t|T|
\)
for every \(t\in[m]\) and every nonempty coalition \(T\subseteq N\).
The condition for \(t=1\) holds automatically.

Fix \(t\in\{2,\ldots,m\}\). Construct a directed network with
source \(s\), sink \(z\), one vertex for every voter, and one
vertex for every candidate in \(W_t\). Add the following edges:
\begin{enumerate}
    \item an edge \(s\to i\) of capacity \(t\) for every
    voter \(i\in N\);
    \item an edge \(i\to x\) of capacity \(nt+1\) whenever
    \(x\in U_i^t\);
    \item an edge \(x\to z\) of capacity \(n\) for every
    candidate \(x\in W_t\).
\end{enumerate}

The total capacity leaving the source is \(nt\), so the maximum
flow is at most \(nt\). We claim that its value is exactly
\(nt\) if and only if
\(
n|U_T^t|\geq t|T|
\qquad\text{for every }T\subseteq N.
\)

To see this, consider a minimum cut. Its capacity is at most
\(nt\), since the cut placing only \(s\) on the source side
has capacity \(nt\). Thus, no edge of capacity \(nt+1\)
can cross the cut. In particular, whenever a voter \(i\)
is on the source side, all candidates in \(U_i^t\) must also
be on the source side.

Let \(T\) be the set of voters on the source side. We may
assume that precisely the candidates in \(U_T^t\) are also
on the source side, since moving any additional candidate
to the sink side can only decrease the cut capacity. The
resulting cut has capacity
\(t(n-|T|)+n|U_T^t|\).
Consequently, the minimum-cut value is
\(\min_{T\subseteq N}\left\{t(n-|T|)+n|U_T^t|\right\}\),
where \(U_\emptyset^t=\emptyset\).

By the max-flow min-cut theorem, the maximum-flow value
equals \(nt\) if and only if
\[t(n-|T|)+n|U_T^t|\geq nt \qquad\text{for every }T\subseteq N.\]
 This is equivalent to
\(n|U_T^t|\geq t|T|\),
which is precisely the local condition.

Thus, running this maximum-flow computation for each
\(t=2,\ldots,m\) verifies Robust Rank-PSC$+$.

Moreover, if the maximum-flow value is smaller than \(nt\),
a minimum cut returns a nonempty coalition \(T\) satisfying
\(n|U_T^t|<t|T|\).
Let \(\ell=|U_T^t|\). Taking \(k=t-1\) and \(X=\{c_t\}\),
we obtain
\(|T|>\ell\frac nt\),
while
\(c_t\notin W_{t-1}\)
and
\(|W_{t-1}\cap X|+|D_T(X)\cap W_t|=|U_T^t|-1=\ell-1\).
Thus, \((k,t,T,X,\ell)\) explicitly witnesses a violation
of Robust Rank-PSC$+$.

Each network has at most \(n+m+2\) vertices and
\(nm+n+m\) edges. Hence, the entire verification procedure
requires at most \(m-1\) polynomial-time maximum-flow
computations.
\end{proof}

\subsection{CoNP-Completeness of Verifying Robust Rank-PSC}

Although rankings satisfying robust Rank-PSC can be computed in polynomial time, we show that verifying whether a given ranking satisfies this property is coNP-complete. Our hardness proof proceeds in two steps. We first establish coNP-completeness for robust PSC in committee selection through a reduction from \textsc{Vertex Cover} on tripartite graphs. We then extend the construction to show coNP-completeness for complete rankings.

We begin with the following restriction of \textsc{Vertex Cover}, which remains NP-complete.

\begin{lemma}\label{lem:verification-tripartite-cover}
\textsc{Vertex Cover} is NP-complete on tripartite graphs with a
supplied partition $U\mathbin{\dot\cup}V\mathbin{\dot\cup}Z$, even
when the budget $t$ satisfies
$|Z|\le t\le |U|+|V|+|Z|$.
\end{lemma}
\begin{proof}
Membership in NP is immediate. Given a \textsc{Vertex Cover} instance
$(G,K)$ with $0\le K\le |V(G)|$, replace every edge $uv$ by the path
$u-p_{uv}-q_{uv}-v$. Put the original vertices in $U$, the vertices
$p_{uv}$ in $V$, and the vertices $q_{uv}$ in $Z$, and set
$t=|E(G)|+K$.

An original cover extends by one internal vertex per path.
Conversely, if a new cover selects original vertices $S$ and leaves
$q$ original edges uncovered by $S$, it must select at least
$|E(G)|+q$ internal vertices. A cover of size at most $t$ therefore
has $|S|+q\le K$; adding one endpoint of each uncovered edge gives
an original cover of size at most $K$.
\end{proof}

Using this restriction, we establish the hardness of verifying robust PSC for committees.

\begin{theorem}\label{thm:verification-committee}
Under strict preferences, deciding whether a given committee satisfies
robust PSC is coNP-complete, even when $n=k+1$.
\end{theorem}

We next extend the hardness result from committees to complete rankings, showing that requiring robust PSC at every prefix does not make verification computationally easier.

\begin{theorem}\label{thm:verification-ranking}
Under strict preferences, deciding whether a given complete ranking
satisfies robust Rank-PSC is coNP-complete, even when $n=m+1$.
\end{theorem}

\section{Conclusion}

We studied proportional representation in rankings, with a particular focus on extending representation guarantees beyond perfectly solid coalitions. Starting from the proportional veto core, we introduced candidatewise and setwise notions of proportional veto protection for rankings and established a connection between setwise protection and robust proportionality for almost-solid coalitions. We showed that these guarantees can be satisfied efficiently, providing polynomial-time constructions of rankings satisfying robust Rank-PSC$+$. We also investigated approval-based proportionality for rankings, establishing both positive results and limitations on combining it with solid-coalition representation.

Our results suggest several directions for future research. An immediate open question is whether a committee satisfying both robust PSC and Rank-PJR$+$ can be computed in polynomial time. Our construction establishes their compatibility and runs in polynomial time for every fixed committee size, but its running time has an exponent depending on the committee size. More broadly, our results highlight a fundamental tension between approval-based proportionality and PSC-based guarantees for rankings. A better understanding of this landscape, including which meaningful combinations of the two approaches are achievable, is an interesting direction for future work.

\section*{AI Acknowledgments}

ChatGPT (GPT-6.1 Sol) was extensively used throughout the development of this paper, including in the generation of many of the proofs. However, the proof of \Cref{thm:PSC-Rank-PJR} was primarily driven by an original insight of the authors, who identified the tension between existing voting rules and developed the central idea for reconciling the two proportionality guarantees. ChatGPT subsequently assisted in completing and formalizing the proof. ChatGPT was also used extensively in drafting and polishing the manuscript. The authors have carefully reviewed all arguments and take full responsibility for the correctness of the proofs and the content of the paper.

\bibliographystyle{plainnat}
\bibliography{bibliography}

\appendix

\section{Rank-PVC Guarantees of Veto Procedures}
\label{app:candidatewise-veto}

In this section, we show that the rankings obtained by
running SimultaneousVeto, or SerialVeto with a fixed
round-robin order of voters, satisfy Candidatewise Rank-PVC.
In both cases, candidates are ranked in reverse elimination
order, so that the last candidate eliminated is placed first.

\begin{theorem}
Every ranking returned by SimultaneousVeto satisfies
Candidatewise Rank-PVC.
\end{theorem}

\begin{proof}
Let $\sigma$ be the ranking returned by SimultaneousVeto.
Fix a candidate $a\in C$ and a nonempty coalition
$T\subseteq N$, and let
$r=\operatorname{rank}_{\sigma}(a)$.

Recall that every candidate initially has capacity $n$,
and every voter continuously vetoes their least-preferred
remaining candidate at unit speed. Let $t$ denote the
time at which candidate $a$ is eliminated.

Since candidates are ranked in reverse elimination order,
at least $m-r+1$ candidates, including $a$, have been
eliminated by time $t$. Each eliminated candidate has
received $n$ units of veto weight. Since all voters
jointly veto at rate $n$, we obtain
\[
nt\geq n(m-r+1),
\]
and therefore
\[
t\geq m-r+1.
\]

On the other hand, until time $t$, candidate $a$
remains available. Consequently, every voter in $T$
vetoes only candidates that they rank weakly below $a$.
Thus, all veto weight contributed by voters in $T$
up to time $t$ is assigned to candidates in $N_a(T)$.

Since the voters in $T$ jointly veto at rate $|T|$,
they contribute a total of $|T|t$ units of veto weight.
Moreover, the candidates in $N_a(T)$ have total
initial capacity $n|N_a(T)|$. Hence,
\[
n|N_a(T)|\geq |T|t.
\]
Combining the two inequalities gives
\[
n|N_a(T)|
\geq |T|(m-r+1),
\]
or equivalently,
\[
\frac{|N_a(T)|}{m}
\geq
\frac{|T|}{n}\cdot\frac{m-r+1}{m}.
\]
Since $a$ and $T$ were arbitrary, $\sigma$ satisfies
Candidatewise Rank-PVC.
\end{proof}

We next show that the same guarantee holds for
SerialVeto when voters perform their veto operations
in a fixed round-robin order.

\begin{theorem}
Every ranking returned by SerialVeto with a fixed
round-robin order of voters satisfies Candidatewise
Rank-PVC.
\end{theorem}

\begin{proof}
Let $\sigma$ be the ranking returned by SerialVeto.
Fix a candidate $a\in C$ and a nonempty coalition
$T\subseteq N$, and let
$r=\operatorname{rank}_{\sigma}(a)$.

Recall that every candidate initially has capacity $n$,
and each veto operation decreases the capacity of the
voter's least-preferred remaining candidate by one.
The voters perform their operations in a fixed
round-robin order, repeated for $m$ rounds.

Let $p$ denote the number of veto operations performed
up to and including the elimination of $a$.

Since $a$ is the $(m-r+1)$-st candidate eliminated,
at least $m-r+1$ candidates have exhausted their
capacity by step $p$. Each such candidate requires
$n$ veto operations. Therefore,
\[
p\geq n(m-r+1).
\]

Under the round-robin schedule, every voter performs
at least $\lfloor p/n\rfloor$ veto operations among
the first $p$ operations. Consequently, every voter
in $T$ has performed at least $m-r+1$ veto operations
by the time $a$ is eliminated.

Moreover, candidate $a$ remains available until its
elimination. Thus, whenever a voter in $T$ performs
a veto operation before or at the elimination of $a$,
the voter vetoes a candidate ranked weakly below $a$.
In particular, all these veto operations are directed
toward candidates in $N_a(T)$.

It follows that voters in $T$ have collectively
contributed at least
\[
|T|(m-r+1)
\]
veto operations to candidates in $N_a(T)$.

Since every candidate has initial capacity $n$,
the candidates in $N_a(T)$ can receive at most
$n|N_a(T)|$ veto operations. Therefore,
\[
n|N_a(T)|
\geq |T|(m-r+1).
\]
Rearranging gives
\[
\frac{|N_a(T)|}{m}
\geq
\frac{|T|}{n}\cdot\frac{m-r+1}{m}.
\]

Since $a$ and $T$ were arbitrary, $\sigma$ satisfies
Candidatewise Rank-PVC.
\end{proof}

The round-robin assumption is important for the second
result. It guarantees that every voter has performed at
least $m-r+1$ veto operations by the time the candidate
at position $r$ is eliminated. For an arbitrary veto
order, this need not hold, and SerialVeto does not
necessarily satisfy Candidatewise Rank-PVC, as shown
by the following example.

\begin{example}
Consider an election with three voters and four candidates
$C=\{a,b,c,d\}$, with preferences
\[
\begin{array}{c|l}
\text{Voters} & \text{Preference ranking}\\ \hline
1,2 & d\succ c\succ b\succ a\\
3   & d\succ a\succ c\succ b
\end{array}
\]

Each candidate initially has capacity $3$, and each
voter performs four veto operations. Consider the
following admissible veto order:
\[
(3,3,3,3,1,1,1,1,2,2,2,2).
\]
Under this order, SerialVeto eliminates the candidates
in the order
\[
b,\ a,\ c,\ d.
\]
Thus, the resulting social ranking is
\[
\sigma=(d,c,a,b).
\]

Now consider candidate $a$, which appears at position
$r=3$, and the coalition $T=\{1,2\}$. Both voters in
$T$ rank $a$ last, so
\[
N_a(T)=\{a\}.
\]
Consequently,
\[
\frac{|N_a(T)|}{m}=\frac14,
\]
whereas Candidatewise Rank-PVC requires
\[
\frac{|T|}{n}\cdot\frac{m-r+1}{m}
=
\frac23\cdot\frac24
=
\frac13.
\]
Since
\[
\frac14<\frac13,
\]
the ranking violates Candidatewise Rank-PVC.

\end{example}

Unlike the Candidatewise Rank-PVC guarantee, Setwise Rank-PVC
is not necessarily satisfied by simply ranking candidates
in the reverse order of their elimination under
SerialVeto or SimultaneousVeto. 

\begin{example}
Consider an election with five voters and four candidates
\(C=\{a,b,c,d\}\), with preferences
\[
\begin{array}{c|l}
\text{Voter} & \text{Preference ranking}\\ \hline
1 & b\succ a\succ d\succ c\\
2 & b\succ d\succ c\succ a\\
3 & c\succ a\succ b\succ d\\
4 & a\succ c\succ d\succ b\\
5 & b\succ d\succ c\succ a
\end{array}
\]

Consider first SimultaneousVeto. Initially, each candidate
has weight \(5\), and every voter continuously vetoes
their least-preferred remaining candidate at unit speed.
Candidate \(a\) is eliminated first, at time \(5/2\),
followed by \(c\) at time \(10/3\), and \(d\) at time
\(15/4\), leaving \(b\) as the winner.
The same elimination order is obtained by SerialVeto
under the round-robin schedule \((1,2,3,4,5)\), repeated
for four rounds. Thus, both procedures eliminate the
candidates in the order
$a,\ c,\ d,\ b$. 
Ranking candidates in reverse elimination order gives
$\sigma=(b,d,c,a)$. Now consider the coalition \(T=\{3,4\}\), the candidate
set \(X=\{b,d\}\), and the prefix of size \(q=2\).
We have \(X=W_2(\sigma)\). Moreover, both voters in
\(T\) rank \(a\) and \(c\) above every candidate in
\(X\). Consequently,
$N_X(T)=\{b,d\}$.

It follows that
$|N_X(T)|-|X|+1=2-2+1=1$,
whereas
\[
\frac{|T|}{n}(m-q+1)
=\frac25\cdot3=\frac65.
\]
Therefore,
\[
|N_X(T)|-|X|+1
<
\frac{|T|}{n}(m-q+1),
\]
violating Setwise Rank-PVC.

\end{example}

\section{Omitted Proofs} \label{app:omitted-proofs}

\subsection*{Proof of~\Cref{prop:running-time}}
\begin{proof}

It suffices to show that each safe-purchase test can be implemented
within the claimed running time. The main difficulty is that the
reserve constraints range over exponentially many voter coalitions
and candidate sets. We show that these constraints admit an efficient
separation oracle.

Fix a prospective committee $W'$ and a nonnegative vector $z$
of remaining voter budgets. Recall that a claim $(T,X)$ is active
only if its entitlement is positive and has not yet been met.
In particular,
\(
|\disag_T(X)|
<
\left\lfloor\sfrac{|T|}{q}\right\rfloor
\le k.
\)
Thus, every active claim has a disagreement set
$D=\disag_T(X)$ of size less than $k$.
We enumerate all such sets $D\subseteq C$, of which there are
at most $m^{O(k)}$.

Fix one such set $D$. If $\disag_T(X)=D$, then deleting
$D$ from the voters' rankings makes $X$ a top prefix of
every voter in $T$. Consequently, it suffices to consider
the at most $nm$ nonempty prefix sets appearing in the
rankings restricted to $C\setminus D$.

Fix such a pair $(D,X)$. For each voter $i$, let
\(
D_i=\disag_{\{i\}}(X).
\)
A coalition $T$ satisfies $\disag_T(X)=D$ if and only if
$D_i\subseteq D$ for every $i\in T$ and
$\bigcup_{i\in T}D_i=D$.

We can find the minimum remaining budget among all such
coalitions of each cardinality by dynamic programming.
For every $s\in\{0,\ldots,n\}$ and $M\subseteq D$,
compute
\[
F(s,M)=
\min\left\{
\sum_{i\in T}z_i:
|T|=s,\ 
\bigcup_{i\in T}D_i=M,
D_i\subseteq D\ \forall i\in T
\right\},
\]
where the minimum of an empty family is $+\infty$.
Processing eligible voters one at a time, we either include
or exclude each voter, updating the coalition size,
disagreement union, and total budget.
Since $|D|<k$, this takes $O(n^2 2^k)$ time.

For a coalition of size $s$ with disagreement set $D$,
its robust PSC entitlement is
\[
h_s=\min\left\{
|X|,
\max\left\{0,\left\lfloor\frac{s}{q}\right\rfloor-|D|\right\}
\right\}.
\]
If $h_s>|W'\cap X|$, the corresponding reserve constraint is
\[
\sum_{i\in T}z_i
\ge
q\left(h_s-|W'\cap X|+|D\setminus W'|\right).
\]
Since the right-hand side depends only on $s$, $D$, $X$,
and $W'$, this constraint holds for every eligible coalition
of size $s$ if and only if
\[
F(s,D)\ge
q\left(h_s-|W'\cap X|+|D\setminus W'|\right).
\]
Thus, the dynamic program either verifies all constraints
associated with $(D,X)$ or returns a coalition witnessing
a violation. Every active claim is considered, so this
gives an exact separation oracle running in
\(
m^{O(k)}\operatorname{poly}(n,m)
\)
time.

Now consider a candidate $c$ at approval depth $r$.
A safe purchase requires payments $p_i$ satisfying
\[
0\le p_i\le b_i,\qquad
\sum_{i\in N}p_i=q,\qquad
p_i=0\quad\text{if }c\notin A_i^r,
\]
together with all reserve constraints for
$W'=W\cup\{c\}$ and $z_i=b_i-p_i$.
These are linear inequalities in the payment variables.
The separation oracle above, combined with the explicit
payment constraints, therefore allows us to decide
feasibility and find a safe payment vector using
polynomial-time linear programming with separation.

\end{proof}

\subsection*{Proof of~\Cref{thm:robust-rank-psc-plus-local}}
\begin{proof}
Suppose first that \(\sigma\) satisfies Robust Rank-PSC$+$.
The condition for \(t=1\) holds automatically, since
\(U_T^1=\{c_1\}\). Suppose that the claimed inequality fails
for some \(t\geq2\) and some nonempty coalition \(T\). Set
\(\ell=|U_T^t| =1+|D_T(\{c_t\})\cap W_t|\).
The failure implies
\(|T|>\ell\frac nt\).
Applying Robust Rank-PSC$+$ with \(k=t-1\) and
\(X=\{c_t\}\), we obtain
\(|D_T(\{c_t\})\cap W_t|\geq\ell\),
since \(c_t\notin W_{t-1}\). This contradicts
\(|D_T(\{c_t\})\cap W_t|=\ell-1\).

Conversely, suppose that
\(n|U_T^t|\geq t|T|\)
holds for every \(t\) and every nonempty coalition \(T\).
Suppose, towards a contradiction, that Robust Rank-PSC$+$
is violated. Then there exist \(1\leq k\leq t\leq m\),
a nonempty coalition \(T\subseteq N\), a nonempty set
\(X\subseteq W_t\), and an integer \(\ell\geq1\) such that
\(|T|>\ell\frac{n}{k+1}\),
while
\(X\nsubseteq W_k\) and \(|W_k\cap X|+|D_T(X)\cap W_t|<\ell\).

Let \(r\) be the smallest index greater than \(k\) for
which \(c_r\in X\). Such an index exists, and by its choice,
\(|W_r\cap X|=|W_k\cap X|+1\).
Moreover, every candidate in \(U_T^r\setminus X\)
belongs to \(D_T(X)\cap W_t\). Indeed, such a candidate is
ranked above \(c_r\in X\) by some voter in \(T\).
Consequently,
\[
\begin{aligned}
|U_T^r|
&\leq |W_r\cap X|+|D_T(X)\cap W_t|\\
&=|W_k\cap X|+1+|D_T(X)\cap W_t|\\
&\leq\ell.
\end{aligned}
\]
Applying the assumed local condition at \(r\), we obtain
\[
\frac{|T|}{n}
\leq\frac{|U_T^r|}{r}
\leq\frac{\ell}{r}
\leq\frac{\ell}{k+1},
\]
contradicting the assumed coalition-size condition.
Therefore, \(\sigma\) satisfies Robust Rank-PSC$+$.
\end{proof}

\subsection*{Proof of~\Cref{thm:verification-committee}}
\begin{proof}

We start with the following lemma. 
\begin{lemma}[Normalized witnesses]\label{lem:verification-normalized}
A size-$k$ committee $W$ violates robust PSC if and only
if some nonempty $T\subseteq N$ and $X\subseteq C\setminus W$ satisfy
\begin{equation}\label{eq:verification-normalized}
(k+1)|T|>n\bigl(|D_T(X)|+1\bigr).
\end{equation}
\end{lemma}
\begin{proof}
A violation with sets $T,Y$ and demand $\ell$ has
$Y\nsubseteq W$ and
$\ell\ge |Y\cap W|+|D_T(Y)|+1$.
For $X=Y\setminus W\ne\emptyset$, we have
\[
|D_T(X)|\le |D_T(Y)|+|Y\cap W|,
\]
so the same coalition satisfies
\eqref{eq:verification-normalized}.
Conversely, \eqref{eq:verification-normalized} gives a violation
with demand $\ell=|D_T(X)|+1$.
\end{proof}

Membership in coNP follows from
Lemma~\ref{lem:verification-normalized}.
For hardness, take a tripartite graph with partition
$U\mathbin{\dot\cup}V\mathbin{\dot\cup}Z$ and budget $t$ as in
Lemma~\ref{lem:verification-tripartite-cover}.
Let $E_{UZ}$ be its edges between $U$ and $Z$, and set
\[
R=t+1,\qquad n_0=(2+|E_{UZ}|)R+|V|,
\qquad d=n_0-t-2.
\]
Create disjoint candidate sets
\[
B=U\cup\{a\},\qquad |D|=d,\qquad
H=Z\cup Q,\quad |H|=t+1,
\]
where $a$ and the members of $D,Q$ are new candidates. The designated
committee is $W_0=D\cup H$, of size $k_0=n_0-1$.

All voters rank $D$ first in the same order. There are:
\begin{enumerate}
\item $R$ copies of each of two reference orders $D\succ B\succ H$,
      with opposite orders on $B$ and a common order on $H$;
\item for each $uz\in E_{UZ}$, $R$ voters with order
      $D\succ B\setminus\{u\}\succ z\succ u\succ H\setminus\{z\}$;
\item for each $v\in V$, one voter with order
      $D\succ N_Z(v)\succ B\setminus N_U(v)
      \succ H\setminus N_Z(v)\succ N_U(v)$,
      where $N_U(v),N_Z(v)$ are its neighbors in $U,Z$.
\end{enumerate}
Complete unspecified blocks in arbitrary strict orders.

There are $n_0$ voters. Since $k_0+1=n_0=d+t+2$, the normalized
violation condition is equivalent to the following budget inequality:
\begin{equation}\label{eq:verification-budget}
\underbrace{n_0-|T|}_{\text{omitted voters}}
+\underbrace{|D_T(X)|-d}_{\text{additional defects}}
\le t,
\qquad \emptyset\ne X\subseteq B.
\end{equation}
We show that this holds exactly when the graph has a cover of size
at most $t$.

Given such a cover $S_U\cup S_V\cup S_Z$, choose
$X=\{a\}\cup(U\setminus S_U)$ and include in $T$ every voter except
the vertex voters corresponding to $S_V$.
The displayed orders give
\[
D_T(X)\subseteq D\cup S_U\cup S_Z.
\]
Indeed, an included vertex voter has all its neighbors in
$S_U\cup S_Z$; and an edge voter can introduce its $z$ as a defect
only when its $u$ is outside the cover, forcing $z\in S_Z$.
Thus the left-hand side of \eqref{eq:verification-budget} is at most
$|S_V|+|S_U|+|S_Z|\le t$.

Conversely, suppose \eqref{eq:verification-budget} holds.
Every voter ranks $D$ above $X$, so $|D_T(X)|\ge d$.
At most $t$ voters are therefore omitted, ensuring that $T$ contains
a voter of every reference and edge type. Define
\[
S_U=U\setminus X,\qquad
S_V=\{v\in V:\text{its voter is outside }T\},\qquad
S_Z=Z\cap D_T(X).
\]
The opposite reference orders imply $B\setminus X\subseteq D_T(X)$,
so
\[
|S_U|+|S_V|+|S_Z|
\le |D_T(X)|-d+n_0-|T|\le t.
\]
These sets cover every edge: an uncovered $uz$ would have $u\in X$
and hence $z\in S_Z$ by its edge voter; an uncovered $vz$ would
have its vertex voter in $T$, again forcing $z\in S_Z$.
Finally, an uncovered $uv$ would have $u\in X$ and its vertex voter
in $T$, making all $t+1$ members of $H$ defects, contrary to
\eqref{eq:verification-budget}.
This proves the equivalence. The construction has polynomial size.
\end{proof}

\subsection*{Proof of~\Cref{thm:verification-ranking}}
\begin{proof}
A violating prefix and normalized witness certify nonmembership.
For hardness, retain the committee construction above, put $b=|B|$,
and set
\[
n=(b+1)n_0+1,\qquad |F|=n-b-n_0,
\]
where $F$ consists of new candidates that all original voters rank
last in a common order. Propose the social ranking
\[
\sigma:\quad D\succ H\succ F\succ B,
\]
using the common order on $D$, the reference voters' order on $H$,
and the common order on $F$. Add $n-n_0$ voters reporting $\sigma$.
Then $m=n-1$.

First, any original violation $T,\emptyset\ne X\subseteq B$ persists
at the prefix of size $p_*=n-b-1$. The new candidates introduce no
defect for $T,X$, and $b|T|\le bn_0<n$ gives
\[
\frac{p_*+1}{n}|T|
=|T|-\frac{b|T|}{n}
>|T|-1
\ge |D_T(X)|+1.
\]

Conversely, let $T,X$ be a normalized violation at any prefix
$W_p(\sigma)$. We use one observation: \emph{no voter in $T$ can rank
every member of $W_p(\sigma)$ above any member of $X$}. Otherwise
$|D_T(X)|\ge p$, contradicting
$(p+1)|T|>n(|D_T(X)|+1)$.

The observation excludes all added voters from $T$. It also excludes
any candidate of $D$ from $X$, since $D$ is a common initial block
with the same order for all voters. Consequently, all of $D$ is
counted as defects. As $p+1\le n$, the violation implies
\[
|T|\ge |D_T(X)|+2\ge d+2=n_0-t.
\]
Thus at most $t$ original voters are omitted, and $T$ contains a
reference voter. If $X$ contained an unselected candidate of
$H\cup F$, the prefix would be contained in $D\cup H\cup F$,
where this reference voter agrees with $\sigma$, contradicting the
same observation. Hence $X\subseteq B$.
For these original voters and candidates, $F$ adds no defects, so
$|T|\ge |D_T(X)|+2$ is exactly an original committee violation.
All padding is polynomial, completing the reduction.
\end{proof}

\end{document}